\RequirePackage{fixltx2e}
\documentclass[12pt,a4paper]{iopart}
\pdfoutput=1
\usepackage[utf8]{inputenc}
\usepackage[unicode=true,bookmarksopen=true,colorlinks=true,urlcolor=blue,anchorcolor=blue,citecolor=blue,filecolor=blue,linkcolor=blue,menucolor=blue,pagecolor=blue,linktocpage=true,pdfa=true]{hyperref}

\usepackage{graphicx}

\makeatletter
\providecommand*\@nameundef[1]{\expandafter\let\csname #1\endcsname\@undefined}
\@nameundef{equation*}
\@nameundef{endequation*}
\makeatother
\usepackage{amsmath}
\usepackage{amssymb}
\usepackage{amsfonts}
\usepackage{mathtools}
\usepackage{lmodern}
\usepackage[T1]{fontenc}
\usepackage{bbm}
\DeclareMathAlphabet{\mathbfi}{OML}{cmm}{b}{it}

\let\originalleft\left
\let\originalright\right
\renewcommand{\left}{\mathopen{}\mathclose\bgroup\originalleft}
\renewcommand{\right}{\aftergroup\egroup\originalright}

\makeatletter
\newenvironment{equations}[1][]{\subequations\ifx\relax#1\relax\else\label{#1}\fi\align\ignorespaces}{\endalign\ignorespacesafterend\endsubequations}
\def\@spliteq#1{\begin{equation}\begin{split}#1\end{split}\end{equation}}
\def\splitequation{\collect@body\@spliteq}

\makeatother

\renewcommand{\vec}[1]{{\ifnum9<1#1\mathbf{#1}\else\ifcat\noexpand#1\relax\boldsymbol{#1}\else\mathbfi{#1}\fi\fi}}
\newcommand{\mathe}{\mathrm{e}}
\newcommand{\mathi}{\mathrm{i}}
\let\oldre\Re
\let\oldim\Im
\renewcommand{\Re}{\oldre\mathfrak{e}\,}
\renewcommand{\Im}{\oldim\mathfrak{m}\,}
\newcommand{\total}{\mathop{}\!\mathrm{d}}
\newcommand{\laplace}{\mathop{}\!\bigtriangleup}
\newcommand{\abs}[1]{{\left\lvert{#1}\right\rvert}}
\newcommand{\sgn}{\operatorname{sgn}}

\newcommand{\eqend}[1]{\,#1}
\newcommand{\bigo}[1]{\mathcal{O}\left({#1}\right)}

\newcommand{\expect}[1]{\left\langle{#1}\right\rangle}

\newcommand{\hankel}[1]{\mathop{}\!\mathrm{H}^{(#1)}}
\newcommand{\hypergeom}[2]{\,{}_{#1}\mathrm{F}_{#2}}

\usepackage[numbers,sort&compress]{natbib}
\allowdisplaybreaks
\begin{document}

\title{Propagators for gauge-invariant observables in cosmology}

\author{Markus B. Fr{\"o}b${}^\dagger$, William C. C. Lima${}^\ddag$}
\address{${}^\dagger$Department of Mathematics, University of York,\\\leavevmode\hphantom{${}^\dagger$}Heslington, York, YO10 5DD, United Kingdom}
\address{${}^\ddag$Centro de Ci{\^e}ncias Naturais e Humanas, Universidade Federal do ABC,\\\leavevmode\hphantom{${}^\ddag$}Avenida dos Estados 5001, 09210-580, Santo Andr{\'e}, S{\~a}o Paulo, Brazil}
\ead{\href{mailto:mbf503@york.ac.uk}{mbf503@york.ac.uk}}
\ead{\href{mailto:william.lima@ufabc.edu.br}{william.lima@ufabc.edu.br}}

\begin{abstract}
We make a proposal for gauge-invariant observables in perturbative quantum gravity in cosmological spacetimes, building on the recent work of Brunetti et al. [\href{http://dx.doi.org/10.1007/JHEP08(2016)032}{\emph{JHEP} {\bf 08} (2016) 032}]. These observables are relational, and are obtained by evaluating the field operator in a field-dependent coordinate system. We show that it is possible to define this coordinate system such that the non-localities inherent in any higher-order observable in quantum gravity are causal, i.e., the value of the gauge-invariant observable at a point $x$ only depends on the metric and inflation perturbations in the past light cone of $x$. We then construct propagators for the metric and inflaton perturbations in a gauge adapted to that coordinate system, which simplifies the calculation of loop corrections, and give explicit expressions for relevant cases: matter- and radiation-dominated eras and slow-roll inflation.

\end{abstract}

% 04.62.+v	Quantum fields in curved spacetime
% 04.60.-m	Quantum gravity
% 04.60.Bc	Phenomenology of quantum gravity
% 11.10.Lm	Nonlinear or nonlocal theories and models
% 11.15.-q	Gauge field theories
\submitto{CQG}

\maketitle

\section{Introduction}
\label{sec_introduction}

The theory of inflation~\cite{guth1981,sato1981,linde1982,albrechtsteinhardt1982} is the most successful theoretical explanation for the early history of our universe, with observational evidence provided by the anisotropies in the cosmic microwave background (CMB)~\cite{planck2015a,planck2015b,planck2015c}. According to this theory, the anisotropies in the CMB were sourced by quantum fluctuations of the metric, which arise in treating gravity as an effective quantum field theory~\cite{burgess2004}. This approach is known as perturbative quantum gravity and makes predictions valid at energy scales well below the Planck scale. It is therefore the only approach to quantum gravity which has been observationally tested, at least in the linear regime.

Extending the theory to second and higher order, where it becomes nonlinear, one faces a severe obstacle: the construction of diffeomorphism-invariant observables. In perturbative quantum gravity, the diffeomorphism invariance of the underlying gravitational theory translates into a gauge symmetry for the metric perturbations (calling any metric perturbation ``graviton'' for short, and not only the transverse traceless part). In contrast to other gauge theories, such as Yang--Mills theory, this gauge symmetry does not act in an internal space but changes the position of the field itself. As a consequence, local fields (i.e., defined at a point of the background spacetime) cannot be gauge invariant. At linear order it is still possible to find local gauge-invariant observables, such as the linearised Riemann tensor for a flat-space background or the linearised Weyl tensor for conformally flat backgrounds\footnote{In fact, at linear order it is always possible to construct a complete set of local and gauge-invariant observables starting from a certain invariant characterisation of the background spacetime, see~\cite{canepadappiaggikhavkine2017} for the characterisation and~\cite{froebhackhiguchi2017} for the (on-shell) construction of such a set for cosmological backgrounds.}. However, in general this is impossible already at second order, as shown in~\cite{torre1993,giddingsmarolfhartle2006,khavkine2015} at various levels of mathematical sophistication.

Various approaches have been devised to deal with this problem:
\begin{itemize}
\item Instead of the bare field operators, one ``dresses'' them with a graviton cloud in such a way that the resulting composite operator is invariant~\cite{waresaotomeakhoury2013,donnellygiddings2015,donnellygiddings2016}.
\item In order to take into account the quantum fluctuations of the metric, one defines the distance between points in correlation functions not using the background metric, but with perturbed geodesics~\cite{mandelstam1962,mandelstam1968,teitelboim1993,woodardthesis,tsamiswoodard1992,modanese1994,urakawatanaka2009,urakawatanaka2010,khavkine2012,bongakhavkine2014,froeb2017a}.
\item One employs the so-called relational observables, which are obtained by considering the field operator at the point where another field has a given value~\cite{giddingsmarolfhartle2006,khavkine2015,marolf2015,brunettifredenhagenrejzner2016}, instead of at a point of the background spacetime. This approach goes back a long way~\cite{geheniaudebever1956a,geheniau1956,debever1956a,debever1956b,geheniaudebever1956b,komar1958,bergmannkomar1960,bergmann1961}, see~\cite{tambornino2012} for a recent review. In general, this amounts to taking scalars constructed from various background fields as configuration-dependent coordinates, and therefore needs a sufficiently generic background spacetime since one needs to be able to differentiate points by the values of these scalars. Alternatively, one could add the scalar fields by hand to the theory (e.g., the famous Brown-Kucha{\v r} dust~\cite{brownkuchar1995}), but this changes the physical content of the theory.
\end{itemize}
In this article, we will adopt the last proposal, where one is immediately faced with the problem of extending the concept also to highly symmetric spacetimes such as cosmological ones. In a single-field inflationary model, the expansion of the background Friedmann--Lema{\^\i}tre--Robertson--Walker (FLRW) spacetime is driven by a scalar field (the inflaton) that only depends on time. All scalars that can be constructed from the background metric and inflaton then also only depend on time, and it is impossible to distinguish points on an equal-time hypersurface by the values of these scalars. This obstacle has been overcome only recently~\cite{brunettietal2016}.

The article is organised as follows: section~\ref{sec_construction} explains how to construct invariant observables in perturbative quantum gravity in detail, with two different (but related) concrete proposals for cosmology. In section~\ref{sec_propagator} we construct propagators for the graviton and inflaton in a gauge adapted to the previous proposals for a general inflaton potential, expressing them in terms of three scalar propagators which depend on the concrete model. We also show there that the second proposal satisfies the conditions needed for a rigorous treatment of higher-order loop corrections. In sections~\ref{sec_epsconst} and~\ref{sec_slowroll} we solve the differential equations for these scalar propagators in two cases relevant in cosmology, namely constant slow-roll parameter $\epsilon$ (which covers matter- and radiation-dominated eras) and slow-roll inflation. We conclude with section~\ref{sec_discussion}, and some technical computations are done in the appendices. We use the `+++' convention of~\cite{mtw_book}, work in $n$ spacetime dimensions, and set $c = \hbar = 1$ and $\kappa^2 = 16 \pi G_\text{N}$.

\section{Construction of invariant observables}
\label{sec_construction}

As explained in the introduction, the main problem in the construction of relational observables around a cosmological background spacetime is the high symmetry of the latter. This obstacle has been overcome only recently~\cite{brunettietal2016}\footnote{We note that for the specific case of the local expansion rate and possible backreaction in inflation, many proposals have been put forward; see, e.g., Refs.~\cite{geshnizjanibrandenberger2002,abramowoodard2002,geshnizjanibrandenberger2005,gasperinimarozziveneziano2010,marozzivacca2012,tsamiswoodard2013,miaotsamiswoodard2017}. We return to this important observable in the discussion.}, and their solution is as follows: Consider a spatially flat FLRW spacetime with background line element
\begin{equation}
\total s^2 = g_{\mu\nu} \total x^\mu \total x^\nu = a^2(\eta) \left( - \total \eta^2 + \total \vec{x}^2 \right) \eqend{,}
\end{equation}
where $\eta$ is conformal time and $a(\eta)$ the scale factor, and a scalar field $\phi$ (the inflaton). Requiring $\phi$ to solve the Klein--Gordon equation with potential $V(\phi)$ in this background, and the FLRW spacetime itself to be a solution of Einstein's equation with the scalar stress tensor as source, we obtain the Friedmann equations
\begin{equations}[eom_background_phi2]
\kappa^2 V(\phi) &= 2 (n-2) (n-1-\epsilon) H^2 \eqend{,} \\
\kappa^2 (\phi')^2 &= 2 (n-2) \epsilon H^2 a^2 \eqend{,}
\end{equations}
where a prime denotes a derivative with respect to conformal time. The Hubble parameter $H$ and the first two slow-roll parameters $\epsilon$ and $\delta$ are defined from the scale factor as
\begin{equation}
\label{H_and_epsilon_def}
H \equiv \frac{a'}{a^2} \eqend{,} \qquad \epsilon \equiv - \frac{H'}{H^2 a} \eqend{,} \qquad \delta \equiv \frac{\epsilon'}{2 H a \epsilon} \eqend{,}
\end{equation}
and are related to the widely used Hubble slow-roll parameters $\epsilon_H$ and $\eta_H$ as~\cite{liddleparsonsbarrow1994}
\begin{equation}
\epsilon = \epsilon_H \eqend{,} \qquad \delta = \epsilon_H - \eta_H \eqend{.}
\end{equation}

We now perturb the above system according to
\begin{equation}
\label{metric_inflaton_perturbation}
g_{\mu\nu} \to \tilde{g}_{\mu\nu} = a^2 \left( \eta_{\mu\nu} + \kappa h_{\mu\nu} \right) \eqend{,} \qquad \phi \to \tilde{\phi} = \phi + \kappa \phi^{(1)} \eqend{.}
\end{equation}
The perturbed inflaton field serves nicely as a field-dependent time coordinate:
\begin{equation}
\label{x0_def}
\tilde{X}^{(0)} = \eta(\tilde{\phi}) \eqend{,}
\end{equation}
where $\eta(\phi)$ is obtained by inverting the background relation $\phi(\eta)$ (assuming an appropriately non-degenerate dependence, i.e., $\phi' \neq 0$). In particular, to first order we obtain
\begin{equation}
\label{x0_sol_firstorder}
\tilde{X}^{(0)}_{(0)}(x) = \eta \eqend{,} \qquad \tilde{X}^{(0)}_{(1)}(x) = \frac{\partial \eta(\phi)}{\partial \phi} \phi^{(1)}(x) = \frac{\phi^{(1)}(x)}{\phi'} \eqend{.}
\end{equation}
To obtain the remaining $n-1$ scalar fields, the authors of~\cite{brunettietal2016} observe that in the background spacetime, the spatial coordinates $x^i$ are harmonic with respect to the spatial Laplacian: $\laplace x^i = 0$. The Laplacian transforms as a scalar under spatial rotations and translations, and one can therefore obtain the $\tilde{X}^{(i)}$ by imposing that they are harmonic with respect to the perturbed Laplacian,
\begin{equation}
\label{xi_def}
\tilde{\laplace}_\phi \tilde{X}^{(i)} = 0 \eqend{,}
\end{equation}
and that on the background they reduce to the $x^i$. We will give an explicit solution of this equation (in perturbation theory) in subsection~\ref{sec_construction_elliptic}. As expected, the resulting $\tilde{X}^{(i)}$ are non-local functionals of the metric and inflaton perturbations, and the physical content of the theory is not altered. However, they have an important drawback: the non-locality is non-causal, in the sense that the value of $\tilde{X}^{(i)}$ at a point $x$ depends on the metric and inflaton perturbation at points which are spacelike separated from $x$. From a mathematical point of view, this means that the usual rigorous approach to quantum field theory in curved spacetime~\cite{hollandswald2001,hollandswald2002,hollandswald2005,hollands2008,fredenhagenrejzner2013,hollandswald2015,hack2015,brunettifredenhagenrejzner2016,fredenhagenrejzner2016} is not applicable, and the renormalisability of loop corrections to invariant observables constructed using these coordinates is not guaranteed. From a physical point of view, the invariant observables constructed using these coordinates are influenced by processes at arbitrarily far spacelike separations, which is clearly undesirable (a form of ``action-at-a-distance'').

Of course, this drawback stems from the fact that we are defining the $\tilde{X}^{(i)}$ as solutions to an elliptic equation, which immediately shows a way out. Namely, on the background the spatial coordinates $x^i$ are also harmonic with respect to the d'Alembertian: $\nabla^2 x^i = 0$. The d'Alembertian also transforms as a scalar, and we obtain another set of configuration-dependent coordinates $\tilde{Y}^{(\mu)}$ by using the same time coordinate,
\begin{equation}
\label{y0_def}
\tilde{Y}^{(0)} = \tilde{X}^{(0)} = \eta(\tilde{\phi}) \eqend{,}
\end{equation}
and then imposing that the spatial $\tilde{Y}^{(i)}$ are harmonic with respect to the perturbed d'Alembertian,
\begin{equation}
\label{yi_def}
\tilde{\nabla}^2 \tilde{Y}^{(i)} = 0 \eqend{.}
\end{equation}
The explicit solution for these coordinates is given in subsection~\ref{sec_construction_hyperbolic}, and this time it is possible to choose the solution such that the non-locality is causal, i.e., the $\tilde{Y}^{(i)}$ at a point $x$ only depend on the metric and inflaton perturbation at points which lie in the past light cone of $x$.

\subsection{An elliptic condition}
\label{sec_construction_elliptic}

To determine the explicit form of the $\tilde{X}^{(i)}$, we first need to define the perturbed spatial Laplacian. For this, we define a time-like unit vector from the gradient of the (perturbed) inflaton,
\begin{equation}
\label{time_u_def}
\tilde{u}_\mu \equiv \frac{\tilde{\nabla}_\mu \tilde{\phi}}{\sqrt{ - \tilde{g}^{\mu\nu} \tilde{\nabla}_\mu \tilde{\phi} \tilde{\nabla}_\nu \tilde{\phi}}} \eqend{,}
\end{equation}
which fulfils $\tilde{u}_\mu \tilde{u}^\mu = -1$, and the induced metric on the hypersurfaces of constant $\tilde{\phi}$
\begin{equation}
\tilde{\gamma}_{\mu\nu} \equiv \tilde{g}_{\mu\nu} + \tilde{u}_\mu \tilde{u}_\nu \eqend{.}
\end{equation}
The perturbed spatial Laplacian is the Laplacian of the constant-inflaton hypersurfaces, which acting on scalar functions has the explicit form
\begin{equation}
\label{spatial_laplacian_def}
\tilde{\laplace}_\phi f = \tilde{\gamma}^\mu_\nu \tilde{\nabla}_\mu \left( \tilde{\gamma}^{\nu\rho} \tilde{\nabla}_\rho f \right) = \tilde{\gamma}^{\mu\nu} \tilde{\nabla}_\mu \tilde{\nabla}_\nu f + \left( \tilde{\nabla}^\mu \tilde{u}_\mu \right) \tilde{u}^\nu \tilde{\nabla}_\nu f \eqend{.}
\end{equation}
On the background, we have $\tilde{u}_\mu^{(0)} = - a \delta^0_\mu$, and it follows that
\begin{equation}
\tilde{\laplace}_\phi^{(0)} f = a^{-2} \laplace f \eqend{.}
\end{equation}
The unique Green's function for $\tilde{\laplace}_\phi^{(0)}$ with vanishing boundary conditions at spatial infinity is $\tilde{G}_\phi^{(0)} \equiv a^2 \laplace^{-1}$, and let us denote by $\tilde{G}_\phi$ the Green's function of $\tilde{\laplace}$. We calculate
\begin{equation}
\tilde{\laplace}_\phi^{(0)} \tilde{G}_\phi^{(0)} = \tilde{\laplace}_\phi \tilde{G}_\phi = \left[ \tilde{\laplace}_\phi^{(0)} + \left( \tilde{\laplace}_\phi - \tilde{\laplace}_\phi^{(0)} \right) \right] \left[ \tilde{G}_\phi^{(0)} + \left( \tilde{G}_\phi - \tilde{G}_\phi^{(0)} \right) \right] \eqend{,}
\end{equation}
from which it follows by convoluting with $\tilde{G}_\phi^{(0)}$ on the left that
\begin{equation}
\tilde{G}_\phi = \tilde{G}_\phi^{(0)} - \tilde{G}_\phi^{(0)} \cdot \left( \tilde{\laplace}_\phi - \tilde{\laplace}_\phi^{(0)} \right) \tilde{G}_\phi \eqend{.}
\end{equation}
Note that the second term on the right-hand side is at least of linear order in perturbations. By repeatedly replacing $\tilde{G}_\phi$ by the right-hand side we obtain a perturbative expansion for $\tilde{G}_\phi$, which reads
\begin{equation}
\tilde{G}_\phi = \sum_{n=0}^\infty \left[ - \tilde{G}_\phi^{(0)} \cdot \left( \tilde{\laplace}_\phi - \tilde{\laplace}_\phi^{(0)} \right) \right] \tilde{G}_\phi^{(0)} \eqend{.}
\end{equation}
The explicit solution of equation~\eqref{xi_def} for the $\tilde{X}^{(i)}$ is then given by
\begin{equation}
\label{xi_sol}
\tilde{X}^{(i)} = \left( 1 - \tilde{G}_\phi \cdot \tilde{\laplace}_\phi \right) x^i = \sum_{n=0}^\infty \left[ - \tilde{G}_\phi^{(0)} \cdot \left( \tilde{\laplace}_\phi - \tilde{\laplace}_\phi^{(0)} \right) \right]^n x^i \eqend{.}
\end{equation}
Applying $\tilde{\laplace}_\phi$, this clearly fulfils equation~\eqref{xi_def}, and the solution is non-trivial because $\tilde{G}_\phi \cdot \tilde{\laplace}_\phi$ is the identity only on functions which vanish at spatial infinity, which is not the case for the $x_i$. In particular, to first order we obtain
\begin{equation}
\label{xi_sol_firstorder}
\tilde{X}^{(i)}_{(0)}(x) = x^i \eqend{,} \qquad \tilde{X}^{(i)}_{(1)}(x) = - \int \tilde{G}_\phi^{(0)}(x,y) \tilde{\laplace}_\phi^{(1)}(y) y^i \total^n y \eqend{.}
\end{equation}

\subsection{A hyperbolic condition}
\label{sec_construction_hyperbolic}

The construction of the $\tilde{Y}^{(i)}$ proceeds absolutely analogous to the elliptic case, only substituting the perturbated spatial Laplacian by the perturbed d'Alembertian. However, now the background Green's function $\tilde{G}^{(0)}$ of the background d'Alembertian $\tilde{\nabla}^2_{(0)} = \nabla^2 = a^{-2} \left[ \partial^2 - (n-2) H a \partial_0 \right]$ is not unique. Different choices of that Green's function correspond to different choices of initial conditions, which in turn define different coordinates $\tilde{Y}^{(i)}$. In a recent explicit calculation in the usual flat-space in-out formalism, the Feynman propagator was used to construct a certain field-dependent coordinate system which greatly simplified the graviton loop corrections to a gauge-invariant quantity~\cite{froeb2017b}. However, in general curved spacetimes one is usually interested in the (causal) evolution of the expectation value of observables, for which the retarded Green's function $\tilde{G}_\text{ret}$ is needed, and which corresponds to the initial conditions $\left. \tilde{Y}^{(i)} \right\rvert_{\eta \to -\infty} = x^i$ and $\left. \partial_\eta \tilde{Y}^{(i)} \right\rvert_{\eta \to -\infty} = 0$. The explicit solution of equation~\eqref{yi_def} for the $\tilde{Y}^{(i)}$ is then given by the analogue of equation~\eqref{xi_sol}, namely
\begin{equation}
\label{yi_sol}
\tilde{Y}^{(i)} = \left( 1 - \tilde{G}_\text{ret} \cdot \tilde{\nabla}^2 \right) x^i = \sum_{n=0}^\infty \left[ - \tilde{G}^{(0)}_\text{ret} \cdot \left( \tilde{\nabla}^2 - \tilde{\nabla}^2_{(0)} \right) \right]^n x^i \eqend{.}
\end{equation}
Again, the solution is non-trivial because the $x^i$ do not vanish in the infinite past, and in particular at first order we obtain
\begin{equation}
\label{yi_sol_firstorder}
\tilde{Y}^{(i)}_{(0)}(x) = x^i \eqend{,} \qquad \tilde{Y}^{(i)}_{(1)}(x) = - \int \tilde{G}^{(0)}_\text{ret}(x,y) \tilde{\nabla}^2_{(1)}(y) y^i \total^n y \eqend{.}
\end{equation}

\subsection{Invariant observables}
\label{sec_construction_obs}

Using the field-dependent coordinates defined above, we can now construct invariant observables. For simplicity, we restrict to the case of the $\tilde{X}^{(\mu)}$; the corresponding construction using the $\tilde{Y}^{(\mu)}$ is again completely analogous. We first need to invert the functional relation $\tilde{X}^{(\mu)}(x)$ to obtain the background coordinates $x^\mu$ as functionals of the field-dependent coordinates $\tilde{X}^{(\mu)}$. To first order, this is easy to do:
\begin{equation}
\label{xmu_inverse_firstorder}
x^\mu = \tilde{X}^{(\mu)} - \kappa \tilde{X}^{(\mu)}_{(1)}(x) + \ldots = \tilde{X}^{(\mu)} - \kappa \tilde{X}^{(\mu)}_{(1)}(\tilde{X}) + \ldots \eqend{,}
\end{equation}
where $\tilde{X}^{(i)}_{(1)}$ is given in equation~\eqref{xi_sol_firstorder}, and $\tilde{X}^{(0)}_{(1)}$ in equation~\eqref{x0_sol_firstorder}. For a scalar field $S$, the corresponding invariant observable $\mathcal{S}$ is obtained by evaluating $S$ at the point $x^\mu$, holding the $\tilde{X}^{(\mu)}$ fixed. Naturally, $S$ itself will have an expansion in terms of metric and inflaton perturbations, given by $S = S_{(0)} + \kappa S_{(1)} + \ldots$, and we have to expand both and collect terms of the same order. To first order, this results in
\begin{equation}
\label{scalar_inv_firstorder}
\mathcal{S}_{(0)} = S_{(0)} \eqend{,} \qquad \mathcal{S}_{(1)} = S_{(1)} - \tilde{X}^{(\mu)}_{(1)} \partial_\mu S_{(0)} \eqend{,}
\end{equation}
where all quantities are now evaluated at $\tilde{X}$. Since the change of a scalar field under infinitesimal diffeomorphisms parametrized by a vector field $\xi^\mu$ is given by
\begin{equation}
\label{scalar_gaugetrafo}
\delta_\xi S = \xi^\rho \partial_\rho S \quad\Rightarrow\quad \delta_\xi S_{(0)} = 0 \eqend{,} \quad \delta_\xi S_{(1)} = \xi^\rho \partial_\rho S_{(0)} \eqend{,}
\end{equation}
and the configuration-dependent coordinates also transform as scalars,
\begin{equation}
\delta_\xi \tilde{X}^{(\mu)}_{(1)} = \xi^\rho \partial_\rho \tilde{X}^{(\mu)}_{(0)} = \xi^\mu \eqend{,}
\end{equation}
we verify that $\mathcal{S}$ is indeed invariant to first order,
\begin{equation}
\delta_\xi \left( \mathcal{S}_{(0)} + \kappa \mathcal{S}_{(1)} \right) = 0 + \ldots \eqend{.}
\end{equation}
We see that the change $x^\mu \to \tilde{X}^{(\mu)}$ is a field-dependent diffeomorphism, which has the effect of compensating for the explicit gauge transformation of fields by including the transformation of the metric and inflaton perturbations.

Note that the $\tilde{X}$ are now just labels (as the coordinates $x^\mu$ had been before); in particular, one should not replace the $\tilde{X}^{(\mu)}$ by their definitions~\eqref{x0_def} or~\eqref{xi_def}, which would only give back the original scalar field $S(x)$. In fact, once the explicit expressions~\eqref{scalar_inv_firstorder} for the invariant observable are obtained, one can forget about their origin, and call $\tilde{X}$ again $x$.

For higher-spin fields, one has also to include the Jacobian arising from the diffeomorphism. For example, a invariant vector field $\mathcal{V}^\mu$ is obtained as
\begin{equation}
\mathcal{V}^\mu = \frac{\partial \tilde{X}^{(\mu)}}{\partial x^\rho} V^\rho(x) = \left( \frac{\partial x^\rho}{\partial \tilde{X}^{(\mu)}} \right)^{-1} V^\rho(x) \eqend{,}
\end{equation}
where the derivative is taken of the functional relation~\eqref{xmu_inverse_firstorder}, and again the $\tilde{X}$ are held fixed. To first order, we calculate
\begin{equation}
\label{vector_inv_firstorder}
\mathcal{V}^\mu_{(0)} = V^\mu_{(0)} \eqend{,} \qquad \mathcal{V}^\mu_{(1)} = V^\mu_{(1)} - \tilde{X}^{(\rho)}_{(1)} \partial_\rho V^\mu_{(0)} + \left( \partial_\rho \tilde{X}^{(\mu)}_{(1)} \right) V^\rho_{(0)} \eqend{,}
\end{equation}
and using that the change of a vector under infinitesimal diffeomorphisms reads
\begin{equation}
\delta_\xi V^\mu = \xi^\rho \partial_\rho V^\mu - V^\rho \partial_\rho \xi^\mu \quad\Rightarrow\quad \delta_\xi V^\mu_{(0)} = 0 \eqend{,} \quad \delta_\xi V^\mu_{(1)} = \xi^\rho \partial_\rho V^\mu_{(0)} - V^\rho_{(0)} \partial_\rho \xi^\mu \eqend{,}
\end{equation}
it is again straightforward to check that $\mathcal{V}^\mu$ is indeed invariant to first order. The generalisation to higher orders in perturbation theory is also straightforward but somewhat lengthy, such that we refrain from writing the explicit expressions down.

\section{Propagators using suitable gauge conditions}
\label{sec_propagator}

Since the observables constructed with the configuration-dependent coordinates are invariant under gauge transformations, we can compute their correlation functions in any convenient gauge. A suitable gauge is obviously one which makes the first-order coordinate corrections $\tilde{X}^{(\mu)}_{(1)}$ or $\tilde{Y}^{(\mu)}_{(1)}$ vanish, as the amount of terms one needs to compute is decreased substantially. For $\tilde{X}^{(0)}$ (and $\tilde{Y}^{(0)}$) this follows directly from the expansion~\eqref{x0_sol_firstorder}, namely we need to impose $\phi^{(1)} = 0$. For the spatial coordinates, we need to impose $\tilde{\laplace}_\phi^{(1)} x^i = 0$ in order to make $\tilde{X}^{(i)}_{(1)}$~\eqref{xi_sol_firstorder} vanish, while the condition for $\tilde{Y}^{(i)}_{(1)}$~\eqref{yi_sol_firstorder} is $\tilde{\nabla}^2 x^i = 0$. To obtain an explicit expression in terms of the metric perturbation, we expand these operators to first order in metric perturbations [using the definition~\eqref{spatial_laplacian_def}]:
%\tilde{\gamma}^{\mu\nu} = a^{-2} \bar{\eta}^{\mu\nu} - a^{-2} \bar{\delta}^\mu_\rho \bar{\delta}^\nu_\sigma h^{\rho\sigma} - \frac{2}{\phi'} a^{-2} \delta^{(\mu}_0 \bar{\partial}^{\nu)} \phi^{(1)}
% \tilde{\Gamma}^\alpha_{\mu\nu} &= H a \left( 2 \delta^0_{(\mu} \delta^\alpha_{\nu)} + \delta_0^\alpha \eta_{\mu\nu} \right) + \partial_{(\mu} h^\alpha_{\nu)} - \frac{1}{2} \partial^\alpha h_{\mu\nu} + H a \delta_0^\alpha h_{\mu\nu} + H a \eta_{\mu\nu} h^{0\alpha}
\begin{equations}
%\tilde{u}_\mu^{(0)} &= - a \delta^0_\mu \eqend{,} \\
\tilde{u}_\mu^{(1)} &= \frac{a}{2} \delta^0_\mu h_{00} - \frac{a}{\phi'} \bar{\partial}_\mu \phi^{(1)} \eqend{,} \\
%\tilde{\laplace}^{(0)} f &= a^{-2} \laplace f \eqend{,} \\
\begin{split}
\tilde{\laplace}_\phi^{(1)} f &= - a^{-2} \left[ h^{ij} \partial_i \partial_j f + \frac{2}{\phi'} \partial^i \phi^{(1)} \partial_i f' + \frac{1}{\phi'} \laplace \phi^{(1)} f' \right] \\
&\quad- a^{-2} \left[ \partial^k h_{ki} - \frac{1}{2} \partial_i h^k_k + (n-3) \frac{H a}{\phi'} \partial_i \phi^{(1)} \right] \partial^i f \eqend{,}
\end{split} \\
%\tilde{\nabla}^2_{(0)} f &= a^{-2} \left[ \partial^2 f - (n-2) H a f' \right] \eqend{,} \\
\tilde{\nabla}^2_{(1)} f &= - a^{-2} h^{\mu\nu} \partial_\mu \partial_\nu f - a^{-2} \left[ \partial_\nu h^{\mu\nu} - \frac{1}{2} \partial^\mu h + (n-2) H a h^{0\mu} \right] \partial_\mu f \eqend{.}
\end{equations}
Acting on the spatial coordinates, it follows that
\begin{equations}
\tilde{\laplace}_\phi^{(1)} x^i &= - a^{-2} \left[ \partial_k h^{ki} - \frac{1}{2} \partial^i h^k_k + (n-3) \frac{H a}{\phi'} \partial^i \phi^{(1)} \right] \eqend{,} \\
\tilde{\nabla}^2_{(1)} x^i &= - a^{-2} \left[ \partial_\nu h^{i\nu} - \frac{1}{2} \partial^i h + (n-2) H a h^{0i} \right] \eqend{.}
\end{equations}
To impose the vanishing of the right-hand sides even inside time-ordered products (and thus the Feynman propagator in the free theory), we add a Lagrange multiplier (or auxiliary field) term to the action:
\begin{equation}
\label{action_gf}
S_\text{GF} = - \int B_\mu \left[ g_{(0)}^{\mu\nu} u^{(0)}_\nu \phi^{(1)} + \gamma^{(0)}{}^\mu_\nu D^{(1)} x^\nu \right] \sqrt{-g} \total^n x \eqend{,}
\end{equation}
where $D = \tilde{\laplace}_\phi$ for the $\tilde{X}^{(\mu)}$, and $D = \tilde{\nabla}^2$ for the $\tilde{Y}^{(\mu)}$.

To actually determine the propagator in these gauges, it is useful to decompose the metric perturbations according to their transformation under rotations and translations of the spatial hypersurfaces. This decomposition reads~\cite{froeb2014}
\begin{equations}[h_decomp]
h_{00} &= S + 2 X_0' + 2 H a X_0 \eqend{,} \\
h_{0k} &= V_k + X_k' + \partial_k X_0 \eqend{,} \\
h_{kl} &= H_{kl} + \delta_{kl} \Sigma + 2 \partial_{(k} X_{l)} - 2 H a \delta_{kl} X_0 \eqend{,} \\
\phi^{(1)} &= \frac{\phi'}{2 H a} \left( Q + \Sigma - 2 H a X_0 \right) \eqend{,}
\end{equations}
where $S$, $\Sigma$ and $Q$ are scalars, $V_k$ is a transverse vector ($\partial^k V_k = 0$) and $H_{kl}$ is a symmetric transverse traceless tensor ($\partial^k H_{kl} = 0 = \delta^{kl} H_{kl}$). Under a gauge transformation with parameter $\xi_\mu$ the metric and inflaton perturbation change according to
\begin{equation}
\delta_\xi h_{\mu\nu} = \partial_\mu \xi_\nu + \partial_\nu \xi_\mu - 2 H a \eta_{\mu\nu} \xi_0 \eqend{,} \qquad \delta_\xi \phi^{(1)} = - \xi_0 \phi' \eqend{,}
\end{equation}
and $S$, $\Sigma$, $Q$, $V_k$ and $H_{kl}$ are invariant, while the change of $X_\mu$ is given by $\delta_\xi X_\mu = \xi_\mu$. The tensor $H_{kl}$ is the proper graviton (with two polarisations in four dimensions), and $Q$ is the Mukhanov-Sasaki variable~\cite{mukhanovfeldmanbrandenberger1992}. Using this decomposition, the quadratic part of the gravitational action is given by~\cite{froeb2014}
\begin{splitequation}
\label{action_grav}
S_\text{G} &= \frac{1}{4} \int \left[ H^{kl\prime} H_{kl}' + H^{kl} \laplace H_{kl} - 2 V^k \laplace V_k \right] a^{n-2} \total^n x \\
&\quad+ \frac{n-2}{4} \int \left[ \epsilon \left( Q' \right)^2 + \epsilon Q \laplace Q + 2 T \left( \laplace \Sigma + \epsilon H a Q' \right) - (n-1-\epsilon) H^2 a^2 T^2 \right] a^{n-2} \total^n x \eqend{,}
\end{splitequation}
where
\begin{equation}
T \equiv S + \frac{\Sigma'}{H a} + \epsilon (Q+\Sigma) \eqend{.}
\end{equation}

\subsection{A gauge for the elliptic condition}
\label{sec_propagator_elliptic}

Using the decomposition~\eqref{h_decomp}, the gauge-fixing term~\eqref{action_gf} for $D = \tilde{\laplace}_\phi$ reads
\begin{splitequation}
\label{action_gf_elliptic}
S_\text{GF} &= \int \left[ - \frac{\phi'}{2 H} B_0 \left( Q + \Sigma - 2 H a X_0 \right) + B_i \left( \laplace X^i + \frac{n-3}{2} \partial^i Q \right) \right] a^{n-2} \total^n x \\
&= \int \left[ B_i^\text{T} \laplace X^i_\text{T} - C_0 Y_0 - B Y \right] a^{n-2} \total^n x \eqend{,}
\end{splitequation}
where we defined
\begin{splitequation}
\label{elliptic_redefinition}
B &\equiv \partial^i B_i \eqend{,} \qquad B_i^\text{T} \equiv \Pi_i^j B_j \eqend{,} \qquad X \equiv \partial^i X_i \eqend{,} \qquad X_i^\text{T} \equiv \Pi_i^j X_j \eqend{,} \\
Y_0 &\equiv X_0 - \frac{Q + \Sigma}{2 H a} \eqend{,} \qquad Y \equiv \partial^i X_i + \frac{n-3}{2} Q \eqend{,} \qquad C_0 \equiv - B_0 a \phi' \eqend{,}
\end{splitequation}
with the transverse projector
\begin{equation}
\label{projector_1_def}
\Pi_{ij} \equiv \delta_{ij} - \frac{\partial_i \partial_j}{\laplace} \eqend{.}
\end{equation}
The expression for the gravitational action can be further simplified by using
\begin{equation}
U \equiv \laplace \Sigma + \epsilon H a Q' \eqend{,} \qquad V \equiv T - \frac{U}{(n-1-\epsilon) H^2 a^2} \eqend{,}
\end{equation}
such that equation~\eqref{action_grav} reduces to
\begin{splitequation}
\label{action_grav_elliptic}
S_\text{G} &= \frac{1}{4} \int H^{kl} P_\text{H} H_{kl} \total^n x + \frac{1}{2} \int Q P_\text{Q} Q \total^n x - \frac{1}{2} \int V^k P_\text{V} V_k \total^n x \\
&\quad+ \frac{n-2}{4} \int \frac{a^{n-4}}{(n-1-\epsilon) H^2} U^2 \total^n x - \frac{n-2}{4} \int (n-1-\epsilon) H^2 a^n V^2 \total^n x \eqend{,}
\end{splitequation}
where we defined the symmetric differential operators
\begin{equations}[diffops]
P_\text{H} &\equiv a^{n-2} \left( \partial^2 - (n-2) H a \partial_0 \right) \eqend{,} \\
P_\text{Q} &\equiv \frac{n-2}{2} a^{n-2} \epsilon \left( \partial^2 - (n-2+2\delta) H a \partial_0 \right) \eqend{,} \\
P_\text{V} &\equiv a^{n-2} \laplace \eqend{.}
\end{equations}

Next, let us assume that we know the propagators corresponding to the differential operators~\eqref{diffops}, which fulfil
\begin{equation}
\label{propagators_ghq}
P_{\text{H}/\text{Q}} \, G^\text{F}_{\text{H}/\text{Q}}(x,x') = \delta^n(x-x') \eqend{,}
\end{equation}
and where the superscript F stands for ``Feynman''. These can be written explicitly in time-ordered form:
\begin{equation}
\label{feynman_propagator}
G^\text{F}_{\text{H}/\text{Q}}(x,x') = \Theta(\eta-\eta') G^+_{\text{H}/\text{Q}}(x,x') + \Theta(\eta'-\eta) G^+_{\text{H}/\text{Q}}(x',x) \eqend{,}
\end{equation}
where the superscript $+$ denotes the Wightman function, and the corresponding retarded propagators are
\begin{equation}
\label{retarded_propagator}
G^\text{ret}_{\text{H}/\text{Q}}(x,x') \equiv G^\text{F}_{\text{H}/\text{Q}}(x,x') - G^+_{\text{H}/\text{Q}}(x',x) = \Theta(\eta-\eta') \left[ G^+_{\text{H}/\text{Q}}(x,x') - G^+_{\text{H}/\text{Q}}(x',x) \right] \eqend{.}
\end{equation}
Since $P_\text{H}$ and $P_\text{Q}$ are hyperbolic operators, they have unique retarded Green's functions with the proper support in the past light cone.\footnote{Since the field operators commute for spacelike separations, we have $G^+_{\text{H}/\text{Q}}(x',x) = G^+_{\text{H}/\text{Q}}(x,x')$ in this case and the term in brackets vanishes outside the light cone.} The situation is different for $P_\text{V}$, which is an elliptic operator. In fact, since the only regular solution of $\laplace f = 0$ (with vanishing boundary conditions at spatial infinity) is $f = 0$, there is no Wightman function for $P_\text{V}$, which would have to satisfy the homogeneous equation $P_\text{V} \, G^+_\text{V}(x,x') = 0$. One can of course invert $P_\text{V}$ and obtain a ``Feynman propagator'', which reads
\begin{equation}
\label{feynman_laplacian}
G^\text{F}_{\text{V}}(x,x') = a^{2-n} \laplace^{-1} \delta^n(x-x') = - a^{2-n} \delta(\eta-\eta') \frac{\Gamma\left( \frac{n-3}{2} \right)}{4 \pi^\frac{n-1}{2} \abs{\vec{x}-\vec{x}'}^{n-3}} \eqend{,}
\end{equation}
where the second equality is valid for $n \geq 4$. The corresponding ``retarded propagator'', defined in analogy to the hyperbolic case~\eqref{retarded_propagator}, is equal to the ``Feynman propagator'' since the Wightman function vanishes. It is clear from the explicit expression~\eqref{feynman_laplacian} that it does not have the proper support, since it does not vanish when $x$ and $x'$ are spacelike separated. We will see at the end of this section that the problem persists for the correlation functions of the metric and inflaton perturbation and the auxiliary field $B_\mu$.

Since the gravitational action~\eqref{action_grav_elliptic} is diagonal in field space, we can easily invert it to obtain the time-ordered correlation functions
\begin{equations}[correlator_elliptic_grav]
- \mathi \expect{ \mathcal{T} H_{ij}(x) H_{kl}(x') } &= \left( 2 \Pi_{i(k} \Pi_{l)j} - \frac{2}{n-2} \Pi_{ij} \Pi_{kl} \right) G^\text{F}_\text{H}(x,x') \eqend{,} \\
- \mathi \expect{ \mathcal{T} V_i(x) V_j(x') } &= - \Pi_{ij} \, a^{2-n} \laplace^{-1} \delta^n(x-x') \eqend{,} \\
- \mathi \expect{ \mathcal{T} Q(x) Q(x') } &= G^\text{F}_\text{Q}(x,x') \eqend{,} \\
- \mathi \expect{ \mathcal{T} U(x) U(x') } &= \frac{2}{n-2} (n-1-\epsilon) H^2 a^{4-n} \delta^n(x-x') \eqend{,} \\
- \mathi \expect{ \mathcal{T} V(x) V(x') } &= - \frac{2}{n-2} \frac{1}{(n-1-\epsilon) H^2 a^n} \delta^n(x-x') \eqend{,}
\end{equations}
with all other correlators between these variables vanishing, and where the projectors ensure transversality (for $H_{kl}$ and $V_k$) and tracelessness (for $H_{kl}$). The corresponding Wightman functions are given by the same expressions, replacing $G^\text{F} \to G^+$ and $\delta^n \to 0$, and the retarded correlation functions by replacing $G^\text{F} \to G^\text{ret}$ (and leaving the $\delta^n$ untouched). Since the Wightman and retarded correlation functions are always obtained in this way, in the following we will only give the expressions for the time-ordered correlation functions. Finding the propagators for the other variables is a bit more tricky since the gauge-fixing action~\eqref{action_gf_elliptic} is not diagonal. To calculate the inverse, we use the following trick: Assume an action of the form $\int A P B \total^n x$, where $A$ and $B$ are fields and $P$ a symmetric differential operator with propagator $G^\text{F}$. We add the action $\alpha/2 \int B P B \total^n x$ and shift the fields to obtain
\begin{equation}
\int A P B \total^n x + \frac{\alpha}{2} \int B P B \total^n x = - \frac{1}{2 \alpha} \int A P A \total^n x + \frac{\alpha}{2} \int \tilde{B} P \tilde{B} \total^n x
\end{equation}
with $\tilde{B} = B + \frac{1}{\alpha} A$. This can now be inverted to get
\begin{equations}
- \mathi \expect{ \mathcal{T} A(x) A(x') } &= - \alpha G^\text{F}(x,x') \eqend{,} \\
- \mathi \expect{ \mathcal{T} A(x) \tilde{B}(x') } &= - \mathi \expect{ \mathcal{T} \tilde{B}(x) A(x') } = 0 \eqend{,} \\
- \mathi \expect{ \mathcal{T} \tilde{B}(x) \tilde{B}(x') } &= \frac{1}{\alpha} G^\text{F}(x,x') \eqend{,}
\end{equations}
and undoing the shift we obtain
\begin{equations}
- \mathi \expect{ \mathcal{T} A(x) A(x') } &= - \alpha G^\text{F}(x,x') \eqend{,} \\
- \mathi \expect{ \mathcal{T} A(x) B(x') } &= - \mathi \expect{ \mathcal{T} B(x) A(x') } = G^\text{F}(x,x') \eqend{,} \\
- \mathi \expect{ \mathcal{T} B(x) B(x') } &= 0 \eqend{.}
\end{equations}
The original propagators are now obtained in the limit $\alpha \to 0$, and we see that only the cross term between $A$ and $B$ is non-vanishing. It follows that
\begin{equations}[correlator_elliptic_gauge]
- \mathi \expect{ \mathcal{T} B_i^\text{T}(x) X_j^\text{T}(x') } &= - \mathi \expect{ \mathcal{T} X_j^\text{T}(x) B_i^\text{T}(x') } = a^{2-n} \laplace^{-1} \delta^n(x-x') \eqend{,} \\
- \mathi \expect{ \mathcal{T} C_0(x) Y_0(x') } &= - \mathi \expect{ \mathcal{T} Y_0(x) C_0(x') } = - a^{2-n} \delta^n(x-x') \eqend{,} \\
- \mathi \expect{ \mathcal{T} B(x) Y(x') } &= - \mathi \expect{ \mathcal{T} Y(x) B(x') } = - a^{2-n} \delta^n(x-x') \eqend{,}
\end{equations}
and all other correlators vanishing.

To obtain the correlation functions of the metric and inflaton perturbation and the auxiliary field, we have to express them in terms of the variables~\eqref{elliptic_redefinition} [using also the decomposition~\eqref{h_decomp}], which results in
\begin{splitequation}
h_{00} &= V + \frac{U}{(n-1-\epsilon) H^2 a^2} + \frac{Q'}{H a} + 2 Y_0' + 2 H a Y_0 \eqend{,} \\
h_{0k} &= V_k + X_k^{\text{T}\prime} + \frac{\partial_k}{\laplace} \left( \frac{U + \laplace Q}{2 H a} - \frac{n-3+\epsilon}{2} Q' + Y' + \laplace Y_0 \right) \eqend{,} \\
h_{kl} &= H_{kl} + 2 \partial_{(k} X_{l)}^\text{T} - \frac{\partial_k \partial_l}{\laplace} \left[ (n-3) Q - 2 Y \right] - \delta_{kl} \left( Q + 2 H a Y_0 \right) \eqend{,} \\
\phi^{(1)} &= - \phi' Y_0 \eqend{,} \qquad B_0 = - \frac{C_0}{a \phi'} \eqend{,} \qquad B_i = B_i^\text{T} + \frac{\partial_i}{\laplace} B \eqend{.}
\end{splitequation}
Using the above correlation functions~\eqref{correlator_elliptic_grav} and~\eqref{correlator_elliptic_gauge}, we finally obtain
\begin{equations}[correlator_elliptic_hphi]
- \mathi \left\langle \mathcal{T} \phi^{(1)}(x) \phi^{(1)}(x') \right\rangle &= - \mathi \left\langle \mathcal{T} \phi^{(1)}(x) h_{\mu\nu}(x') \right\rangle = 0 \eqend{,} \\
- \mathi \left\langle \mathcal{T} h_{00}(x) h_{00}(x') \right\rangle &= \frac{1}{(Ha)(\eta) (Ha)(\eta')} \partial_\eta \partial_{\eta'} G^\text{F}_\text{Q}(x,x') \eqend{,} \\
\begin{split}
- \mathi \left\langle \mathcal{T} h_{00}(x) h_{0k}(x') \right\rangle &= \frac{1}{(Ha)(\eta)} \partial_\eta \left[ \frac{n-3+\epsilon(\eta')}{2} \frac{\partial_{\eta'}}{\laplace} - \frac{1}{2 (Ha)(\eta')} \right] \partial_k G^\text{F}_\text{Q}(x,x') \\
&\quad- \frac{a^{2-n}}{(n-2) H a} \frac{\partial_k}{\laplace} \delta^n(x-x') \eqend{,}
\end{split} \\
- \mathi \left\langle \mathcal{T} h_{00}(x) h_{kl}(x') \right\rangle &= - \widehat{\Pi}_{kl} \frac{1}{(Ha)(\eta)} \partial_\eta G^\text{F}_\text{Q}(x,x') \eqend{,} \\
\begin{split}
- \mathi \left\langle \mathcal{T} h_{0i}(x) h_{0k}(x') \right\rangle &= - \partial_i \partial_k \left[ \frac{n-3+\epsilon(\eta)}{2} \frac{\partial_\eta}{\laplace} - \frac{1}{2 (Ha)(\eta)} \right] \\
&\qquad\qquad\times \left[ \frac{n-3+\epsilon(\eta')}{2} \frac{\partial_{\eta'}}{\laplace} - \frac{1}{2 (Ha)(\eta')} \right] G^\text{F}_\text{Q}(x,x') \\
&\quad- a^{2-n} \left[ \delta_{ik} - \frac{n-3+\epsilon}{2 (n-2)} \frac{\partial_i \partial_k}{\laplace} \right] \laplace^{-1} \delta^n(x-x') \eqend{,}
\end{split} \\
- \mathi \left\langle \mathcal{T} h_{0i}(x) h_{kl}(x') \right\rangle &= \left[ \frac{n-3+\epsilon(\eta)}{2} \frac{\partial_\eta}{\laplace} - \frac{1}{2 (Ha)(\eta)} \right] \partial_i \widehat{\Pi}_{kl} \, G^\text{F}_\text{Q}(x,x') \eqend{,} \\
- \mathi \left\langle \mathcal{T} h_{ij}(x) h_{kl}(x') \right\rangle &= \left( 2 \Pi_{i(k} \Pi_{l)j} - \frac{2}{n-2} \Pi_{ij} \Pi_{kl} \right) G^\text{F}_\text{H}(x,x') + \widehat{\Pi}_{ij} \widehat{\Pi}_{kl} \, G^\text{F}_\text{Q}(x,x')
\end{equations}
with the second projector
\begin{equation}
\label{projector_2_def}
\widehat{\Pi}_{ij} \equiv \delta_{ij} + (n-3) \frac{\partial_i \partial_j}{\laplace} \eqend{,}
\end{equation}
and
\begin{equations}[correlator_elliptic_b]
- \mathi \left\langle \mathcal{T} B_0(x) \phi^{(1)}(x') \right\rangle &= - a^{1-n} \delta^n(x-x') \eqend{,} \\
- \mathi \left\langle \mathcal{T} B_0(x) h_{00}(x') \right\rangle &= - \frac{2 a^{1-n}}{\phi'} \left( \partial_\eta - H a \right) \delta^n(x-x') \eqend{,} \\
- \mathi \left\langle \mathcal{T} B_0(x) h_{0k}(x') \right\rangle &= - \frac{a^{1-n}}{\phi'} \partial_k \delta^n(x-x') \eqend{,} \\
- \mathi \left\langle \mathcal{T} B_0(x) h_{kl}(x') \right\rangle &= - \delta_{kl} \frac{2 H a^{2-n}}{\phi'} \delta^n(x-x') \eqend{,} \\
- \mathi \left\langle \mathcal{T} B_i(x) \phi^{(1)}(x') \right\rangle &= 0 \eqend{,} \\
- \mathi \left\langle \mathcal{T} B_i(x) h_{00}(x') \right\rangle &= 0 \eqend{,} \\
- \mathi \left\langle \mathcal{T} B_i(x) h_{0k}(x') \right\rangle &= - \delta_{ik} a^{2-n} \partial_\eta \laplace^{-1} \delta^n(x-x') \eqend{,} \\
- \mathi \left\langle \mathcal{T} B_i(x) h_{kl}(x') \right\rangle &= - 2 a^{2-n} \frac{\delta_{i(k} \partial_{l)}}{\laplace} \delta^n(x-x') \eqend{.}
\end{equations}

If we now form the corresponding retarded Green's functions by subtracting the Wightman function from the Feynman propagator, we obtain results without the proper support (in the past light cone), as explained before. This can be seen very clearly for the auxiliary field, where we have for example
\begin{equation}
- \mathi \left\langle B_i(x) h_{0k}(x') \right\rangle^\text{ret} = - \delta_{ik} a^{2-n} \partial_\eta \laplace^{-1} \delta^n(x-x') \eqend{,}
\end{equation}
since the Wightman function in this case vanishes; and this expression does not vanish for spacelike separated $x$ and $x'$, as follows from the explicit form of $\laplace^{-1} \delta^n(x-x')$~\eqref{feynman_laplacian}. One might think that since the auxiliary field only enforces the gauge condition~\eqref{action_gf}, this does not represent a problem in the quantum theory, but the problem also appears for the correlator of the metric perturbations. Let us define
\begin{equations}[feynman_d_hq]
D^\text{F}_\text{H}(x,x') &\equiv \laplace^{-1} \left[ \partial_\eta \partial_{\eta'} G^\text{F}_\text{H}(x,x') - a^{2-n} \delta^n(x-x') \right] \eqend{,} \\
D^\text{F}_\text{Q}(x,x') &\equiv \laplace^{-1} \left[ \partial_\eta \partial_{\eta'} G^\text{F}_\text{Q}(x,x') - \frac{2 a^{2-n}}{(n-2) \epsilon} \delta^n(x-x') \right] \eqend{,}
\end{equations}
and the corresponding retarded Green's functions by replacing $G^\text{F}$ by $G^\text{ret}$. It is shown in~\ref{appendix_support} that those have the proper support (in the past light cone for the retarded Green's functions), since the Dirac $\delta$ is cancelled when the time derivatives act on the time-ordering $\Theta$ functions. Replacing second mixed time derivatives by the above, we obtain for example
\begin{splitequation}
- \mathi \left\langle h_{00}(x) h_{0k}(x') \right\rangle^\text{ret} &= \frac{1}{(Ha)(\eta)} \left[ \frac{n-3+\epsilon(\eta')}{2} \partial_k D^\text{ret}_\text{Q}(x,x') - \frac{1}{2 (Ha)(\eta')} \partial_k \partial_\eta G^\text{ret}_\text{Q}(x,x') \right] \\
&\quad+ \frac{n-3}{n-2} \frac{a^{1-n}}{H \epsilon} \frac{\partial_k}{\laplace} \delta^n(x-x') \eqend{.} \raisetag{1.5em}
\end{splitequation}
While the two terms in the first line have the proper support (and no extra terms involving a Dirac $\delta$ are generated from a single time derivative, as is also shown in~\ref{appendix_support}), there is a $\laplace^{-1} \delta^n(x-x')$ remaining, which does not vanish for spacelike separated $x$ and $x'$~\eqref{feynman_laplacian}.

\subsection{A gauge for the hyperbolic condition}
\label{sec_propagator_hyperbolic}

In this case, it is not possible to find variables to decouple the gauge-fixing and the gravitational action, and we have to work with the sum $S = S_\text{G} + S_\text{GF}$ from the outset. Using the decomposition~\eqref{h_decomp}, the sum of gravitational action~\eqref{action_grav} and gauge-fixing term~\eqref{action_gf} for $D = \tilde{\nabla}^2$ reads
\begingroup
\setlength{\arraycolsep}{2pt}
\begin{splitequation}
\label{action_hyperbolic}
S &= - \frac{1}{2} \int B_0 \frac{\phi'}{H} \left( Q + \Sigma - 2 H a X_0 \right) a^{n-2} \total^n x + \int B_i \Big[ - V^{i\prime} - (n-2) H a V^i \\
&\qquad\qquad- \frac{n-3}{2} \partial^i \Sigma + \frac{1}{2} \partial^i S + \left( \partial^2 - (n-2) H a \partial_0 \right) X^i \Big] a^{n-2} \total^n x + S_\text{G} \\
&= \frac{1}{4} \int H^{kl} P_\text{H} H_{kl} \total^n x - \frac{1}{2} \int W^i P_\text{V} W_i \total^n x + \int B_0 Y_0 \phi' a^{n-1} \total^n x \\
&\qquad+ \frac{1}{2} \int \begin{pmatrix} Y^i & X^i_\text{T} \end{pmatrix} \laplace \begin{pmatrix} P_\text{H} - P_\text{V} & P_\text{V} \\ P_\text{V} & - P_\text{H} - P_\text{V} \end{pmatrix} \begin{pmatrix} Y_i \\ X_i^\text{T} \end{pmatrix} \total^n x \\
&\qquad+ \frac{n-2}{4} \int \frac{a^{n-4}}{(n-1-\epsilon) H^2} U^2 \total^n x - \frac{n-2}{4} \int (n-1-\epsilon) H^2 a^n V^2 \total^n x \\
&\qquad+ \frac{1}{2} \int \begin{pmatrix} Q & C & Y \end{pmatrix} \begin{pmatrix} P_\text{Q} & \frac{1}{n-2} P_\text{Q} - \frac{\epsilon}{2} P_\text{H} & 0 \\ \frac{1}{n-2} P_\text{Q} - \left( \frac{\epsilon}{2} P_\text{H} \right)^* & - \frac{n-1}{2 (n-2)} P_\text{H} + P_\text{V} + \frac{1}{(n-2)^2} P_\text{Q} & - P_\text{H} \\ 0 & - P_\text{H} & 0 \end{pmatrix} \begin{pmatrix} Q \\ C \\ Y \end{pmatrix} \total^n x \eqend{,}
\end{splitequation}
\endgroup
where we defined
\begin{splitequation}
\label{hyperbolic_redefinition}
B &\equiv \partial^i B_i \eqend{,} \qquad B_i^\text{T} \equiv \Pi_i^j B_j \eqend{,} \qquad X \equiv \partial^i X_i \eqend{,} \qquad X_i^\text{T} \equiv \Pi_i^j X_j \eqend{,} \\
W_i &\equiv V_i - \laplace^{-1} B_i^{\text{T}\prime} \eqend{,} \qquad Y_i \equiv \laplace^{-1} B_i^\text{T} + X_i^\text{T} \eqend{,} \qquad Y_0 \equiv X_0 - \frac{Q + \Sigma}{2 H a} \\
Y &\equiv X - \frac{\epsilon}{2} Q \eqend{,} \qquad U \equiv \laplace \Sigma + \epsilon H a Q' - \frac{1}{n-2} \left[ B + (n-1-\epsilon) H a \laplace^{-1} B' \right] \eqend{,} \\
V &\equiv T - \frac{U}{(n-1-\epsilon) H^2 a^2} - \frac{\laplace^{-1} B'}{(n-2) H a} \eqend{,} \qquad C \equiv \laplace^{-1} B \eqend{,}
\end{splitequation}
and a star denotes the adjoint operator, that is (since $P_\text{H}$ is symmetric)
\begin{equation}
\left( \frac{\epsilon}{2} P_\text{H} \right)^* f = P_\text{H} \left( \frac{\epsilon}{2} f \right) \eqend{.}
\end{equation}
It is straightforward to check that
\begin{equation}
\begin{pmatrix} P_\text{H} - P_\text{V} & P_\text{V} \\ P_\text{V} & - P_\text{H} - P_\text{V} \end{pmatrix} \begin{pmatrix} G^\text{F}_\text{H} + G^\text{F}_2 & G^\text{F}_2 \\ G^\text{F}_2 & - G^\text{F}_\text{H} + G^\text{F}_2 \end{pmatrix}(x,x') = \begin{pmatrix} 1 & 0 \\ 0 & 1 \end{pmatrix} \delta^n(x-x') \eqend{,}
\end{equation}
where $G^\text{F}_2$ is the solution to
\begin{equation}
\label{propagator_g2}
P_\text{H} \, G^\text{F}_2(x,x') = P_\text{V} \, G^\text{F}_\text{H}(x,x') \eqend{.}
\end{equation}
One also checks that
\begin{splitequation}
&\begin{pmatrix} 1 & 0 & 0 \\ 0 & 1 & 0 \\ 0 & 0 & 1 \end{pmatrix} \delta^n(x-x') = \begin{pmatrix} P_\text{Q} & \frac{1}{n-2} P_\text{Q} - \frac{\epsilon}{2} P_\text{H} & 0 \\ \frac{1}{n-2} P_\text{Q} - \left( \frac{\epsilon}{2} P_\text{H} \right)^* & - \frac{n-1}{2 (n-2)} P_\text{H} + P_\text{V} + \frac{1}{(n-2)^2} P_\text{Q} & - P_\text{H} \\ 0 & - P_\text{H} & 0 \end{pmatrix} \\
&\qquad\times \begin{pmatrix} G^\text{F}_\text{Q} & 0 & - \frac{\epsilon(\eta')}{2} G^\text{F}_\text{Q} + \frac{1}{n-2} G^\text{F}_\text{H} \\ 0 & 0 & - G^\text{F}_\text{H} \\ - \frac{\epsilon(\eta)}{2} G^\text{F}_\text{Q} + \frac{1}{n-2} G^\text{F}_\text{H} & - G^\text{F}_\text{H} & \frac{n-1-\epsilon(\eta)-\epsilon(\eta')}{2 (n-2)} G^\text{F}_\text{H} + \frac{\epsilon(\eta) \epsilon(\eta')}{4} G^\text{F}_\text{Q} - G^\text{F}_2 \end{pmatrix}(x,x') \eqend{,}
\end{splitequation}
and we obtain the correlation functions
\begin{equations}[correlator_hyperbolic_grav]
- \mathi \expect{ \mathcal{T} H_{ij}(x) H_{kl}(x') } &= \left( 2 \Pi_{i(k} \Pi_{l)j} - \frac{2}{n-2} \Pi_{ij} \Pi_{kl} \right) G^\text{F}_\text{H}(x,x') \eqend{,} \\
- \mathi \left\langle \mathcal{T} W_i(x) W_j(x') \right\rangle &= - a^{2-n} \Pi_{ij} \laplace^{-1} \delta^n(x-x') \eqend{,} \\
- \mathi \left\langle \mathcal{T} Y_i(x) Y_j(x') \right\rangle &= \Pi_{ij} \laplace^{-1} \left( G^\text{F}_\text{H}(x,x') + G^\text{F}_2(x,x') \right) \eqend{,} \\
- \mathi \left\langle \mathcal{T} Y_i(x) X_j^\text{T}(x') \right\rangle &= \Pi_{ij} \laplace^{-1} G^\text{F}_2(x,x') \eqend{,} \\
- \mathi \left\langle \mathcal{T} X_i^\text{T}(x) X_j^\text{T}(x') \right\rangle &= \Pi_{ij} \laplace^{-1} \left( - G^\text{F}_\text{H}(x,x') + G^\text{F}_2(x,x') \right) \eqend{,} \\
- \mathi \left\langle \mathcal{T} U(x) U(x') \right\rangle &= \frac{2}{n-2} (n-1-\epsilon) H^2 a^{4-n} \delta^n(x-x') \eqend{,} \\
- \mathi \left\langle \mathcal{T} V(x) V(x') \right\rangle &= - \frac{2}{n-2} \frac{1}{(n-1-\epsilon) H^2 a^n} \delta^n(x-x') \eqend{,} \\
- \mathi \left\langle \mathcal{T} Y_0(x) B_0(x') \right\rangle &= \frac{a^{1-n}}{\phi'} \delta^n(x-x') \eqend{,} \\
- \mathi \left\langle \mathcal{T} Q(x) Q(x') \right\rangle &= G^\text{F}_\text{Q}(x,x') \eqend{,} \\
- \mathi \left\langle \mathcal{T} Q(x) Y(x') \right\rangle &= - \frac{\epsilon(\eta')}{2} G^\text{F}_\text{Q}(x,x') + \frac{1}{n-2} G^\text{F}_\text{H}(x,x') \eqend{,} \\
- \mathi \left\langle \mathcal{T} C(x) Y(x') \right\rangle &= - G^\text{F}_\text{H}(x,x') \eqend{,} \\
- \mathi \left\langle \mathcal{T} Y(x) Y(x') \right\rangle &= \frac{n-1-\epsilon(\eta)-\epsilon(\eta')}{2 (n-2)} G^\text{F}_\text{H}(x,x') + \frac{\epsilon(\eta) \epsilon(\eta')}{4} G^\text{F}_\text{Q}(x,x') - G^\text{F}_2(x,x') \eqend{,}
\end{equations}
with all other correlators not related by the exchange of $x$ and $x'$ vanishing.

Reversing the definitions~\eqref{hyperbolic_redefinition} [using the decomposition~\eqref{h_decomp}], we obtain
\begin{splitequation}
h_{00} &= V + \frac{U}{(n-1-\epsilon) H^2 a^2} + \frac{Q'}{H a} + \frac{C'}{(n-2) H a} + 2 Y_0' + 2 H a Y_0 \eqend{,} \\
h_{0k} &= W_k + Y_k' + \frac{\partial_k}{\laplace} \left[ \frac{U}{2 H a} + \epsilon \delta H a Q + \frac{n-1-\epsilon}{2 (n-2)} C' + Y' \right] + \partial_k \left[ Y_0 + \frac{C + (n-2) Q}{2 (n-2) H a} \right] \eqend{,} \\
h_{kl} &= H_{kl} + 2 \partial_{(k} X_{l)}^\text{T} + \frac{\partial_k \partial_l}{\laplace} \left( 2 Y + \epsilon Q \right) - \delta_{kl} \left( Q + 2 H a Y_0 \right) \eqend{,} \\
\phi^{(1)} &= - Y_0 \phi' \eqend{,} \qquad B_0 = B_0 \eqend{,} \qquad B_i = \laplace Y_i - \laplace X_i^\text{T} + \partial_i C \eqend{,} \raisetag{1.5em}
\end{splitequation}
and a long but straightforward calculation using the correlation functions~\eqref{correlator_hyperbolic_grav} gives the correlators of the metric and inflaton perturbation and the auxiliary field. We can simplify the resulting expressions by using equation~\eqref{feynman_d_hq} to replace $\partial_\eta \partial_{\eta'} G^\text{F}_{\text{H}/\text{Q}}$ by $D^\text{F}_{\text{H}/\text{Q}}$ when it is necessary to cancel Dirac $\delta$'s, and using also
\begin{equation}
\label{feynman_d_2}
D^\text{F}_2(x,x') \equiv \frac{\partial_\eta \partial_{\eta'}}{\laplace} G^\text{F}_2(x,x') \eqend{.}
\end{equation}
This leads to
\begin{equations}[correlator_hyperbolic_hphi]
- \mathi \left\langle \mathcal{T} \phi^{(1)}(x) \phi^{(1)}(x') \right\rangle &= - \mathi \left\langle \mathcal{T} \phi^{(1)}(x) h_{\mu\nu}(x') \right\rangle = 0 \eqend{,} \\
- \mathi \left\langle \mathcal{T} h_{00}(x) h_{00}(x') \right\rangle &= \frac{1}{(Ha)(\eta) (Ha)(\eta')} \partial_\eta \partial_{\eta'} G^\text{F}_\text{Q}(x,x') \eqend{,} \\
- \mathi \left\langle \mathcal{T} h_{00}(x) h_{0k}(x') \right\rangle &= \frac{\epsilon(\eta')}{2 (Ha)(\eta)} \partial_k D^\text{F}_\text{Q}(x,x') - \frac{1}{2 (Ha)(\eta) (Ha)(\eta')} \partial_\eta \partial_k G^\text{F}_\text{Q}(x,x') \eqend{,} \\
- \mathi \left\langle \mathcal{T} h_{00}(x) h_{kl}(x') \right\rangle &= - \delta_{kl} \frac{1}{(Ha)(\eta)} \partial_\eta G^\text{F}_\text{Q}(x,x') \eqend{,} \\
\begin{split}
- \mathi \left\langle \mathcal{T} h_{0i}(x) h_{0k}(x') \right\rangle &= \Pi_{ik} \left[ D^\text{F}_\text{H}(x,x') + D^\text{F}_2(x,x') \right] + \frac{\partial_i \partial_k}{\laplace} \bigg[ \frac{n-1}{2 (n-2)} D^\text{F}_\text{H}(x,x') \\
&\qquad\quad+ D^\text{F}_2(x,x') + \frac{(\epsilon H a)(\eta) \partial_\eta + (\epsilon H a)(\eta') \partial_{\eta'} - \laplace}{4 (H a)(\eta) (H a)(\eta')} G^\text{F}_\text{Q}(x,x') \\
&\qquad\quad- \frac{\epsilon(\eta) \epsilon(\eta')}{4} D^\text{F}_\text{Q}(x,x') \bigg] \eqend{,}
\end{split} \\
\begin{split}
- \mathi \left\langle \mathcal{T} h_{0i}(x) h_{kl}(x') \right\rangle &= - 2 \frac{\delta_{i(k} \partial_{l)}}{\laplace} \partial_\eta G^\text{F}_2(x,x') \\
&\quad- \delta_{kl} \frac{\partial_i}{\laplace} \left[ \frac{1}{n-2} \partial_\eta G^\text{F}_\text{H}(x,x') - \left[ \frac{\epsilon(\eta)}{2} \partial_\eta - \frac{\laplace}{2 (H a)(\eta)} \right] G^\text{F}_\text{Q}(x,x') \right] \eqend{,}
\end{split} \\
\begin{split}
- \mathi \left\langle \mathcal{T} h_{ij}(x) h_{kl}(x') \right\rangle &= \left( 2 \delta_{i(k} \delta_{l)j} - \frac{2}{n-2} \delta_{ij} \delta_{kl} \right) G^\text{F}_\text{H}(x,x') + \delta_{ij} \delta_{kl} G^\text{F}_\text{Q}(x,x') \\
&\quad- 4 \frac{\partial_{(i} \delta_{j)(k} \partial_{l)}}{\laplace} G^\text{F}_2(x,x')
\end{split}
\end{equations}
and
\begin{equations}[correlator_hyperbolic_b]
- \mathi \left\langle \mathcal{T} B_0(x) \phi^{(1)}(x') \right\rangle &= - a^{1-n} \delta^n(x-x') \eqend{,} \\
- \mathi \left\langle \mathcal{T} B_0(x) h_{00}(x') \right\rangle &= - \frac{2}{\phi'} a^{1-n} \left( \partial_\eta - H a \right) \delta^n(x-x') \eqend{,} \\
- \mathi \left\langle \mathcal{T} B_0(x) h_{0k}(x') \right\rangle &= - \frac{a^{1-n}}{\phi'} \partial_k \delta^n(x-x') \eqend{,} \\
- \mathi \left\langle \mathcal{T} B_0(x) h_{kl}(x') \right\rangle &= - 2 \delta_{kl} \frac{H a^{2-n}}{\phi'} \delta^n(x-x') \eqend{,} \\
- \mathi \left\langle \mathcal{T} B_i(x) \phi^{(1)}(x') \right\rangle &= 0 \eqend{,} \\
- \mathi \left\langle \mathcal{T} B_i(x) h_{00}(x') \right\rangle &= 0 \eqend{,} \\
- \mathi \left\langle \mathcal{T} B_i(x) h_{0k}(x') \right\rangle &= \delta_{ik} \partial_{\eta'} G^\text{F}_\text{H}(x,x') \eqend{,} \\
- \mathi \left\langle \mathcal{T} B_i(x) h_{kl}(x') \right\rangle &= - 2 \delta_{i(k} \partial_{l)} G^\text{F}_\text{H}(x,x') \eqend{.}
\end{equations}

This time, there are no terms proportional to $\laplace^{-1} \delta^n(x-x')$ remaining, and the only potentially problematic non-local operators are projection operators, or operators of the form $\partial_i \partial_\eta \laplace^{-1}$. It is shown in~\ref{appendix_support} that these do not enlarge the support of retarded propagators, and also $D_2$ can be shown to have the proper support (again, see~\ref{appendix_support}). Therefore, the corresponding retarded Green's functions have the proper support for the hyperbolic gauge condition.

\section{Propagators for constant \texorpdfstring{$\epsilon$}{\textepsilon}}
\label{sec_epsconst}

For constant slow-roll parameter $\epsilon$, we can integrate the defining relations~\eqref{H_and_epsilon_def} to find explicit forms of the Hubble parameter and the scale factor. This results in
\begin{equation}
\label{epsconst_ha}
\delta = 0 \eqend{,} \qquad H = H_0 a^{-\epsilon} \eqend{,} \qquad a = \left[ - (1-\epsilon) H_0 \eta \right]^{- \frac{1}{1-\epsilon}} \eqend{,}
\end{equation}
and for $\epsilon \to 0$ we recover de~Sitter space. A matter-dominated universe has $\epsilon_\text{mat} = (n-1)/2$, while radiation domination is $\epsilon_\text{rad} = n/2$. The differential operator $P_\text{Q}$~\eqref{diffops} is just a constant multiple of $P_\text{H}$:
\begin{equation}
P_\text{Q} = \frac{n-2}{2} \, \epsilon \, P_\text{H} \eqend{.}
\end{equation}
Therefore, the corresponding propagators are simply a constant multiple of each other:
\begin{equation}
\label{propagator_q_epsconst}
G^\text{F}_\text{Q}(x,x') = \frac{2}{(n-2) \epsilon} G^\text{F}_\text{H}(x,x') \eqend{.}
\end{equation}
Furthermore we calculate
\begin{equation}
\label{phpv_commute}
P_\text{H} \left( \eta \partial_\eta + \eta' \partial_{\eta'} - 2 \mu \right) = \left( \eta \partial_\eta + \eta' \partial_{\eta'} + 1 \right) P_\text{H} - 2 P_\text{V} \eqend{,}
\end{equation}
where the parameter $\mu$ is defined by
\begin{equation}
\label{epsconst_mu_def}
\mu \equiv \frac{n-1-\epsilon}{2 (1-\epsilon)} \eqend{,} \qquad \left( \mu_\text{mat} = - \frac{n-1}{2 (n-3)} \eqend{,} \quad \mu_\text{rad} = - \frac{1}{2} \right)
\end{equation}
and it follows that [using that $P_\text{H} \, G^\text{F}_\text{H}(x,x') = \delta^n(x-x')$~\eqref{propagators_ghq}]
\begin{equation}
\label{eps_const_phghcommute}
P_\text{H} \left( \eta \partial_\eta + \eta' \partial_{\eta'} - 2 \mu \right) G^\text{F}_\text{H}(x,x') = - 2 P_\text{V} \, G^\text{F}_\text{H}(x,x') + \left( \eta \partial_\eta + \eta' \partial_{\eta'} + 1 \right) \delta^n(x-x') \eqend{.}
\end{equation}
Since
\begin{equation}
\eta' \partial_{\eta'} \delta(\eta-\eta') = - \partial_\eta \left[ \eta' \delta(\eta-\eta') \right] = - \delta(\eta-\eta') - \eta \partial_\eta \delta(\eta-\eta') \eqend{,}
\end{equation}
the last term in equation~\eqref{eps_const_phghcommute} vanishes, and comparing with equation~\eqref{propagator_g2} we infer that
\begin{equation}
\label{propagator_2_epsconst}
G^\text{F}_2(x,x') = - \frac{1}{2} \left( \eta \partial_\eta + \eta' \partial_{\eta'} - 2 \mu \right) G^\text{F}_\text{H}(x,x') \eqend{.}
\end{equation}

To determine $G^\text{F}_\text{H}(x,x')$, we write the equation $P_\text{H} \, G^\text{F}_\text{H}(x,x') = \delta^n(x-x')$~\eqref{propagators_ghq} in the form
\begin{equation}
\left[ \partial^2 + \frac{(n-2) (n-2\epsilon)}{4 (1-\epsilon)^2 \eta^2} \right] \left[ \left[ a(\eta) a(\eta') \right]^\frac{n-2}{2} G^\text{F}_\text{H}(x,x') \right] = \delta^n(x-x') \eqend{.}
\end{equation}
The propagator of a scalar of mass $m$ in pure de~Sitter space fulfils
\begin{equation}
\label{propagator_desitter}
\left[ \partial^2 + \frac{(n-2) n H_0^2 - 4 m^2}{4 H_0^2 \eta^2} \right] \left[ \left( \frac{1}{H_0^2 \eta \eta'} \right)^\frac{n-2}{2} G^\text{F,dS}_{m^2}(x,x') \right] = \delta^n(x-x') \eqend{,}
\end{equation}
and we thus have
\begin{equation}
\label{propagator_h_epsconst_in_ds}
G^\text{F}_\text{H}(x,x') = \left[ H_0^2 \eta \eta' a(\eta) a(\eta') \right]^\frac{2-n}{2} G^\text{F,dS}_{M^2}(x,x') = \left[ a(\eta) a(\eta') \right]^{- \frac{n-2}{2} \epsilon} G^\text{F,dS}_{M^2}(x,x')
\end{equation}
with
\begin{equation}
\label{propagator_h_epsconst_mass}
M^2 = - (n-2) \epsilon \frac{2 (n-1) - n \epsilon}{4 (1-\epsilon)^2} H_0^2 \eqend{.}
\end{equation}
We note that depending on the value of $\epsilon$, this may not be positive, and we will see that $M^2 < 0$ leads to infrared (IR) divergences in the natural Bunch--Davies vacuum state.

In spatial Fourier space, the de~Sitter Wightman function in this state is given by (see, e.g., \cite{froebhiguchi2014})
\begin{equation}
\tilde{G}^{+,\text{dS}}_{m^2}(\eta,\eta',\vec{p}) = - \mathi H_0^{n-2} \frac{\pi}{4} (\eta\eta')^\frac{n-1}{2} \hankel1_\nu\left( -\abs{\vec{p}} \eta \right) \hankel2_\nu\left( -\abs{\vec{p}} \eta' \right) \eqend{,}
\end{equation}
where the parameter $\nu$ is related to the mass according to
\begin{equation}
\nu \equiv \sqrt{ \frac{(n-1)^2}{4} - \frac{m^2}{H_0^2} } \eqend{.}
\end{equation}
It follows from equations~\eqref{propagator_h_epsconst_in_ds} and~\eqref{epsconst_ha} that
\begin{equation}
\label{wightman_h_epsconst}
\tilde{G}^+_\text{H}(\eta,\eta',\vec{p}) = - \mathi (1-\epsilon)^\frac{(n-2) \epsilon}{1-\epsilon} H_0^\frac{n-2}{1-\epsilon} \frac{\pi}{4} (\eta\eta')^\mu \hankel1_\mu\left( -\abs{\vec{p}} \eta \right) \hankel2_\mu\left( -\abs{\vec{p}} \eta' \right) \eqend{,}
\end{equation}
where the parameter $\mu$ is given by~\eqref{epsconst_mu_def}. For the function $D^+_\text{H}$~\eqref{feynman_d_hq}, we obtain using Hankel function identities~\cite{dlmf}
\begin{equation}
\label{wightman_dh_epsconst}
\tilde{D}^+_\text{H}(\eta,\eta',\vec{p}) = \frac{\partial_\eta \partial_{\eta'}}{\laplace} \tilde{G}^+_\text{H}(\eta,\eta',\vec{p}) = \mathi (1-\epsilon)^\frac{(n-2) \epsilon}{1-\epsilon} H_0^\frac{n-2}{1-\epsilon} \frac{\pi}{4} (\eta\eta')^\mu \hankel1_{\mu-1}\left( -\abs{\vec{p}} \eta \right) \hankel2_{\mu-1}\left( -\abs{\vec{p}} \eta' \right) \eqend{.}
\end{equation}

To obtain explicit expressions in position space, we recall the integral~\cite{froebhiguchi2014}
\begin{splitequation}
\label{fourier_integral_i}
&2^{n-2} \pi^\frac{n+2}{2} (\eta\eta')^\frac{n-1}{2} \int \hankel1_\nu\left( -\abs{\vec{p}} \eta \right) \hankel2_\nu\left( -\abs{\vec{p}}\eta' \right) \mathe^{\mathi \vec{p} \vec{r}} \frac{\total^{n-1} p}{(2\pi)^{n-1}} \\
&= \frac{\Gamma\left( \frac{n-1}{2}+\nu \right) \Gamma\left( \frac{n-1}{2}-\nu \right)}{\Gamma\left( \frac{n}{2} \right)} \hypergeom{2}{1}\left[ \frac{n-1}{2}+\nu, \frac{n-1}{2}-\nu; \frac{n}{2}; \frac{1+Z}{2} - \mathi 0 \sgn(\eta-\eta') \right]
\end{splitequation}
with
\begin{equation}
Z \equiv 1 - \frac{\vec{r}^2 - (\eta-\eta')^2}{2 \eta \eta'} \eqend{,}
\end{equation}
which is valid whenever the integrand does not diverge too strongly for small $\abs{\vec{p}}$; in the case above, this means $\abs{\Re \nu} < (n-1)/2$. In single-field inflation, we have $\epsilon > 0$~\eqref{eom_background_phi2}, and it is easy to see that $\Re \mu > (n-1)/2$~\eqref{epsconst_mu_def} for all $0 < \epsilon < 1$. Therefore, we can not use this integral for the Wightman function; one possibility is to cut it off for some small $\abs{\vec{p}}$~\cite{tsamiswoodard1994,janssenetal2008}. This is of course the well-known IR problem for massless fields in cosmological spacetimes~\cite{fordparker1977a,fordparker1977b}, and appears also for the Feynman propagator $G^\text{F}_\text{H}$~\eqref{feynman_propagator}. However, we are mostly interested in the retarded Green's function $G^\text{ret}_\text{H}$~\eqref{retarded_propagator}, which is independent of the quantum state, and where these state-dependent IR divergences cancel. Namely, in spatial Fourier space we have
\begin{splitequation}
\tilde{G}^\text{ret}_\text{H}(\eta,\eta',\vec{p}) &= \Theta(\eta-\eta') \left[ \tilde{G}^+_\text{H}(\eta,\eta',\vec{p}) - \tilde{G}^+_\text{H}(\eta',\eta,\vec{p}) \right] \\
&= - \mathi \, \Theta(\eta-\eta') (1-\epsilon)^\frac{(n-2) \epsilon}{1-\epsilon} H_0^\frac{n-2}{1-\epsilon} \frac{\pi}{4} (\eta\eta')^\mu \\
&\qquad\times \left[ \hankel1_\mu\left( -\abs{\vec{p}} \eta \right) \hankel2_\mu\left( -\abs{\vec{p}} \eta' \right) - \hankel1_\mu\left( -\abs{\vec{p}} \eta' \right) \hankel2_\mu\left( -\abs{\vec{p}} \eta \right) \right] \eqend{,}
\end{splitequation}
which is regular as $\abs{\vec{p}} \to 0$. We can therefore use the integral~\eqref{fourier_integral_i} to obtain
\begin{splitequation}
\label{retarded_h_epsconst}
G^\text{ret}_\text{H}(x,x') &= - \mathi \, \Theta(\eta-\eta') \left[ H(\eta) H(\eta') \right]^\frac{n-2}{2} \frac{\Gamma\left( \frac{n-1}{2}+\mu \right) \Gamma\left( \frac{n-1}{2}-\mu \right)}{(4\pi)^\frac{n}{2} \Gamma\left( \frac{n}{2} \right)} \\
&\qquad\times\bigg[ \hypergeom{2}{1}\left[ \frac{n-1}{2}+\mu, \frac{n-1}{2}-\mu; \frac{n}{2}; \frac{1+Z}{2} - \mathi 0 \sgn(\eta-\eta') \right] \\
&\qquad\qquad\qquad- \hypergeom{2}{1}\left[ \frac{n-1}{2}+\mu, \frac{n-1}{2}-\mu; \frac{n}{2}; \frac{1+Z}{2} + \mathi 0 \sgn(\eta-\eta') \right] \bigg] \eqend{,}
\end{splitequation}
the discontinuity of the hypergeometric function across its branch cut. An explicit expression for the discontinuity is calculated in~\ref{appendix_hypergeom}, and the final result~\eqref{appendix_hypergeom_jump_sing} strongly depends on the spacetime dimension $n$. For $n = 4$, we obtain after some simplifications
\begin{splitequation}
G^\text{ret}_\text{H}(x,x') &= - \Theta(\eta-\eta') \frac{H(\eta) H(\eta')}{4 \pi} \\
&\qquad\times\left[ \frac{2-\epsilon}{2 (1-\epsilon)^2} \Theta(Z-1) \hypergeom{2}{1}\left( \frac{3-2\epsilon}{1-\epsilon}, - \frac{\epsilon}{1-\epsilon}; 2; \frac{1-Z}{2} \right) + \delta(Z-1) \right] \eqend{,}
\end{splitequation}
which is of the general Hadamard form~\cite{hadamard1932,dewittbrehme1960,fullingsweenywald1978,baerginouxpfaeffle2007}. For $D^\text{ret}_\text{H}$, we obtain by the same procedure
\begin{splitequation}
D^\text{ret}_\text{H}(x,x') &= - \mathi \, \Theta(\eta-\eta') \left[ H(\eta) H(\eta') \right]^\frac{n-2}{2} \frac{\Gamma\left( \frac{n-3}{2}+\mu \right) \Gamma\left( \frac{n+1}{2}-\mu \right)}{(4\pi)^\frac{n}{2} \Gamma\left( \frac{n}{2} \right)} \\
&\qquad\times\bigg[ \hypergeom{2}{1}\left[ \frac{n-3}{2}+\mu, \frac{n+1}{2}-\mu; \frac{n}{2}; \frac{1+Z}{2} + \mathi 0 \right] \\
&\qquad\qquad\qquad- \hypergeom{2}{1}\left[ \frac{n-3}{2}+\mu, \frac{n+1}{2}-\mu; \frac{n}{2}; \frac{1+Z}{2} - \mathi 0 \right] \bigg]
\end{splitequation}
in the general case, and
\begin{splitequation}
D^\text{ret}_\text{H}(x,x') &= \Theta(\eta-\eta') \frac{H(\eta) H(\eta')}{4 \pi} \\
&\qquad\times\left[ \frac{\epsilon}{2 (1-\epsilon)^2} \Theta(Z-1) \hypergeom{2}{1}\left( \frac{2-\epsilon}{1-\epsilon}, \frac{1-2\epsilon}{1-\epsilon}; 2; \frac{1-Z}{2} \right) + \delta(Z-1) \right]
\end{splitequation}
for $n = 4$.

In the matter-dominated case where $\epsilon = 3/2$ (for $n = 4$), this simplifies further to
\begin{equations}
G^\text{ret}_\text{H}(x,x') &= - \Theta(\eta-\eta') \frac{H(\eta) H(\eta')}{4 \pi} \left[ \Theta(Z-1) + \delta(Z-1) \right] \eqend{,} \\
D^\text{ret}_\text{H}(x,x') &= \Theta(\eta-\eta') \frac{H(\eta) H(\eta')}{4 \pi} \left[ 3 \Theta(Z-1) (Z-1) + \delta(Z-1) \right] \eqend{,}
\end{equations}
and in the radiation-dominated case with $\epsilon = 2$, we obtain
\begin{equations}
G^\text{ret}_\text{H}(x,x') &= - \Theta(\eta-\eta') \frac{H(\eta) H(\eta')}{4 \pi} \delta(Z-1) \eqend{,} \\
D^\text{ret}_\text{H}(x,x') &= \Theta(\eta-\eta') \frac{H(\eta) H(\eta')}{4 \pi} \left[ \Theta(Z-1) + \delta(Z-1) \right] \eqend{.}
\end{equations}

\subsection{The elliptic condition}
\label{sec_epsconst_elliptic}

Expressing the propagators $G^\text{F}_\text{Q}$~\eqref{propagator_q_epsconst} and $G^\text{F}_2$~\eqref{propagator_2_epsconst} in terms of $G^\text{F}_\text{H}$, the correlation functions~\eqref{correlator_elliptic_hphi} simplify. From equation~\eqref{epsconst_ha} it follows that $H a = - [ (1-\epsilon) \eta ]^{-1}$, and we obtain
\begin{equations}[correlator_elliptic_hphi_epsconst]
- \mathi \left\langle \mathcal{T} \phi^{(1)}(x) \phi^{(1)}(x') \right\rangle &= - \mathi \left\langle \mathcal{T} \phi^{(1)}(x) h_{\mu\nu}(x') \right\rangle = 0 \eqend{,} \\
- \mathi \left\langle \mathcal{T} h_{00}(x) h_{00}(x') \right\rangle &= \frac{2 (1-\epsilon)^2}{(n-2) \epsilon} \eta \eta' \partial_\eta \partial_{\eta'} G^\text{F}_\text{H}(x,x') \eqend{,} \\
\begin{split}
- \mathi \left\langle \mathcal{T} h_{00}(x) h_{0k}(x') \right\rangle &= - \frac{1-\epsilon}{(n-2) \epsilon} \eta \partial_\eta \left[ (n-3+\epsilon) \frac{\partial_{\eta'}}{\laplace} + (1-\epsilon) \eta' \right] \partial_k G^\text{F}_\text{H}(x,x') \\
&\quad- \frac{a^{2-n}}{(n-2) H a} \frac{\partial_k}{\laplace} \delta^n(x-x') \eqend{,}
\end{split} \\
- \mathi \left\langle \mathcal{T} h_{00}(x) h_{kl}(x') \right\rangle &= \frac{2 (1-\epsilon)}{(n-2) \epsilon} \widehat{\Pi}_{kl} \, \eta \partial_\eta G^\text{F}_\text{H}(x,x') \eqend{,} \\
\begin{split}
- \mathi \left\langle \mathcal{T} h_{0i}(x) h_{0k}(x') \right\rangle &= - \frac{1}{2 (n-2) \epsilon} \partial_i \partial_k \left[ (n-3+\epsilon) \frac{\partial_\eta}{\laplace} + (1-\epsilon) \eta \right] \\
&\qquad\qquad\times \left[ (n-3+\epsilon) \frac{\partial_{\eta'}}{\laplace} + (1-\epsilon) \eta' \right] G^\text{F}_\text{H}(x,x') \\
&\quad- a^{2-n} \left[ \delta_{ik} - \frac{n-3+\epsilon}{2 (n-2)} \frac{\partial_i \partial_k}{\laplace} \right] \laplace^{-1} \delta^n(x-x') \eqend{,}
\end{split} \\
- \mathi \left\langle \mathcal{T} h_{0i}(x) h_{kl}(x') \right\rangle &= \frac{1}{(n-2) \epsilon} \left[ (n-3+\epsilon) \frac{\partial_\eta}{\laplace} + (1-\epsilon) \eta \right] \partial_i \widehat{\Pi}_{kl} \, G^\text{F}_\text{H}(x,x') \eqend{,} \\
- \mathi \left\langle \mathcal{T} h_{ij}(x) h_{kl}(x') \right\rangle &= \left( 2 \Pi_{i(k} \Pi_{l)j} - \frac{2}{n-2} \Pi_{ij} \Pi_{kl} + \frac{2}{(n-2) \epsilon} \widehat{\Pi}_{ij} \widehat{\Pi}_{kl} \right) G^\text{F}_\text{H}(x,x') \eqend{,}
\end{equations}
where the projection operators $\Pi_{ij}$ and $\widehat{\Pi}_{ij}$ are defined in equations~\eqref{projector_1_def} and~\eqref{projector_2_def}, respectively. As we have seen previously, the inherent non-locality in this gauge is non-causal, namely the would-be ``retarded'' Green's function does not vanish for spacelike separated points. Therefore, this gauge condition is not very useful from a physical point of view, but for completeness we wanted to give the explicit form~\eqref{correlator_elliptic_hphi_epsconst} of the correlation functions.

\subsection{The hyperbolic condition}
\label{sec_epsconst_hyperbolic}

Again, expressing the propagators $G^\text{F}_\text{Q}$~\eqref{propagator_q_epsconst} and $G^\text{F}_2$~\eqref{propagator_2_epsconst} in terms of $G^\text{F}_\text{H}$, and also $D^\text{F}_\text{Q}$~\eqref{feynman_d_hq} and $D^\text{F}_2$~\eqref{feynman_d_2} by $D^\text{F}_\text{H}$, the correlation functions~\eqref{correlator_hyperbolic_hphi} simplify. For $D^\text{F}_2$, the replacement is not trivial, but using the equation satisfied by $G^\text{F}_\text{H}$~\eqref{propagators_ghq} we calculate
\begin{equations}[epsconst_df2_dfh_relation]
D^\text{F}_2(x,x') &= - \frac{1}{2} \left( \eta \partial_\eta + \eta' \partial_{\eta'} - 2 \mu + 2 \right) D^\text{F}_\text{H}(x,x') \eqend{,} \\
\left( \eta \partial_{\eta'} + \eta' \partial_\eta \right) G^\text{F}_\text{H}(x,x') &= \left( \eta \partial_\eta + \eta' \partial_{\eta'} - 4 \mu + 2 \right) D^\text{F}_\text{H}(x,x') \eqend{.}
\end{equations}
Using also the definition of $\mu$~\eqref{epsconst_mu_def}, we obtain
\begin{equations}[correlator_hyperbolic_hphi_epsconst]
- \mathi \left\langle \mathcal{T} \phi^{(1)}(x) \phi^{(1)}(x') \right\rangle &= - \mathi \left\langle \mathcal{T} \phi^{(1)}(x) h_{\mu\nu}(x') \right\rangle = 0 \eqend{,} \\
- \mathi \left\langle \mathcal{T} h_{00}(x) h_{00}(x') \right\rangle &= \frac{2 (1-\epsilon)^2}{(n-2) \epsilon} \eta \eta' \partial_\eta \partial_{\eta'} G^\text{F}_\text{H}(x,x') \eqend{,} \\
- \mathi \left\langle \mathcal{T} h_{00}(x) h_{0k}(x') \right\rangle &= - \frac{(1-\epsilon) \eta}{(n-2)} \partial_k \left[ D^\text{F}_\text{H}(x,x') + \frac{1-\epsilon}{\epsilon} \eta' \partial_\eta G^\text{F}_\text{H}(x,x') \right] \eqend{,} \\
- \mathi \left\langle \mathcal{T} h_{00}(x) h_{kl}(x') \right\rangle &= \delta_{kl} \frac{2 (1-\epsilon)}{(n-2) \epsilon} \eta \partial_\eta G^\text{F}_\text{H}(x,x') \eqend{,} \\
\begin{split}
- \mathi \left\langle \mathcal{T} h_{0i}(x) h_{0k}(x') \right\rangle &= - \frac{1}{2} \left( \delta_{ik} + \frac{1-\epsilon}{n-2} \frac{\partial_i \partial_k}{\laplace} \right) \left( \eta \partial_\eta + \eta' \partial_{\eta'} - 2 \mu \right) D^\text{F}_\text{H}(x,x') \\
&\quad- \frac{(1-\epsilon)^2}{2 (n-2) \epsilon} \eta \eta' \partial_i \partial_k G^\text{F}_\text{H}(x,x') \eqend{,}
\end{split} \\
- \mathi \left\langle \mathcal{T} h_{0i}(x) h_{kl}(x') \right\rangle &= \delta_{i(k} \partial_{l)} \left[ \eta G^\text{F}_\text{H}(x,x') + \eta' D^\text{F}_\text{H}(x,x') \right] + \delta_{kl} \frac{1-\epsilon}{(n-2) \epsilon} \eta \partial_i G^\text{F}_\text{H}(x,x') \eqend{,} \\
\begin{split}
- \mathi \left\langle \mathcal{T} h_{ij}(x) h_{kl}(x') \right\rangle &= \left( 2 \delta_{i(k} \delta_{l)j} + \frac{2 (1-\epsilon)}{(n-2) \epsilon} \delta_{ij} \delta_{kl} \right) G^\text{F}_\text{H}(x,x') \\
&\quad+ 2 \left( \eta \partial_\eta + \eta' \partial_{\eta'} - 2 \mu \right) \frac{\partial_{(i} \delta_{j)(k} \partial_{l)}}{\laplace} G^\text{F}_\text{H}(x,x')
\end{split}
\end{equations}

In contrast to the elliptic condition, here there are no inverse Laplacians left acting on Dirac $\delta$ distributions, and as explained in~\ref{appendix_support} the projection operators $\partial_i \partial_j \laplace^{-1}$ do not increase the support of retarded Green's functions. Therefore, as anticipated the non-locality in this gauge is causal, and no suspicious ``action-at-a-distance'' results.

\section{Propagators for slow-roll inflation}
\label{sec_slowroll}

In the slow-roll approximation, we assume that $\epsilon \ll 1$ and $\abs{\delta} \ll 1$, and only keep terms linear in the small parameters $\epsilon$ and $\delta$ (see, e.g., Refs.~\cite{liddleparsonsbarrow1994,lidseyetal1997}). From the defining relations~\eqref{H_and_epsilon_def} we see that $\epsilon' = \bigo{\epsilon \delta}$, and can thus neglect $\epsilon'$ except when it is multiplied by an inverse power of a small parameter. We assume that the same is true for $\delta'$. Within this approximation, integrating the relations~\eqref{H_and_epsilon_def} we obtain
\begin{equation}
\label{slowroll_ha}
\epsilon = \epsilon_0 a^{2\delta} \eqend{,} \qquad H = H_0 a^{-\epsilon} \eqend{,} \qquad a = \left[ - (1-\epsilon) H_0 \eta \right]^{- \frac{1}{1-\epsilon}} \eqend{,}
\end{equation}
where $\epsilon_0$ and $H_0$ are constant, and in particular we have
\begin{equation}
\label{slowroll_ha_in_eta}
H a = - \frac{1}{(1-\epsilon) \eta}
\end{equation}
as in the constant $\epsilon$ case. In fact, since most relations are identical to the constant $\epsilon$ case, we can restrict ourselves to the end results, only pointing out the essential differences. This approximation is only valid for some limited range of conformal times $\eta$; expanding the expression~\eqref{slowroll_ha} for $\epsilon$ in powers of $\delta$, we obtain
\begin{equation}
\epsilon = \epsilon_0 \left[ 1 + 2 \delta \ln a + 2 \delta^2 \ln^2 a + \bigo{\delta^3} \right] \eqend{.}
\end{equation}
In order to neglect the third and all higher-order terms, we must have $\abs{ \delta \ln a } \ll 1$, and similarly from the expansion of the expression for $H$ we obtain the condition $\abs{ \epsilon \ln a } \ll 1$. That is, the approximation is valid for as long as the logarithm of the scale factor changes much less than $N = 1/\max\left( \abs{\delta}, \epsilon \right)$. In particular, the two times $\eta$ and $\eta'$ appearing in the propagator that we give in the rest of this section must not be more than $N$ e-foldings apart for the given expressions to be valid. If this condition is satisfied, we can in fact assume $\epsilon$ and $\delta$ to be constant, see for example Ref.~\cite{lidseyetal1997}.\footnote{If one is not interested in the coordinate-space expressions, but only in the results in Fourier space, the approximation can be improved by taking $\epsilon$ and $\delta$ constant but different for each mode, namely at horizon crossing where $H a = \abs{\vec{p}}$; see, e.g., Ref.~\cite{lidseyetal1997}. The condition $\abs{ \{ \delta, \epsilon \} \ln a } \ll 1$ is then unnecessary.}

The first propagator $G_\text{H}^\text{F}(x,x')$ is related to the de~Sitter propagator of mass $M$ in the same way as before~\eqref{propagator_h_epsconst_in_ds}, but the mass~\eqref{propagator_h_epsconst_mass} is now expanded to first order in $\epsilon$:
\begin{equation}
\label{propagator_h_slowroll_mass}
M^2 = - \frac{(n-1) (n-2)}{2} \epsilon H_0^2 \eqend{.}
\end{equation}
It follows that the Wightman function is given (in Fourier space) by
\begin{equation}
\tilde{G}^+_\text{H}(\eta,\eta',\vec{p}) = - \mathi \frac{\pi}{4} \left[ H(\eta) H(\eta') \right]^\frac{n-2}{2} (\eta\eta')^\frac{n-1}{2} \hankel1_\mu\left( -\abs{\vec{p}} \eta \right) \hankel2_\mu\left( -\abs{\vec{p}} \eta' \right) \eqend{,}
\end{equation}
where the parameter $\mu$ is
\begin{equation}
\label{slowroll_mu_def}
\mu \equiv \frac{n-1}{2} + \frac{n-2}{2} \epsilon \eqend{,}
\end{equation}
which coincides with the linearisation of the constant $\epsilon$ case~\eqref{epsconst_mu_def}. Using Hankel function identities~\cite{dlmf}, we also obtain
\begin{equation}
\label{wightman_dh_slowroll}
\tilde{D}^+_\text{H}(\eta,\eta',\vec{p}) = \mathi \frac{\pi}{4} \left[ H(\eta) H(\eta') \right]^\frac{n-2}{2} (\eta\eta')^\frac{n-1}{2} \hankel1_{\mu-1}\left( -\abs{\vec{p}} \eta \right) \hankel2_{\mu-1}\left( -\abs{\vec{p}} \eta' \right)
\end{equation}
for the function $D^+_\text{H}$~\eqref{feynman_d_hq}, which is the linearisation of~\eqref{wightman_dh_epsconst}.

In the slow-roll approximation, equation~\eqref{propagators_ghq} for the second propagator $G^\text{F}_\text{Q}(x,x')$ can be cast in the form
\begin{splitequation}
&\left[ \partial^2 + \frac{n (n-2) + 2 (n-1) (n-2) \epsilon + 4 (n-1) \delta}{4 \eta^2} \right] \\
&\qquad\times \left[ \left[ a(\eta) a(\eta') \right]^\frac{n-2}{2} \sqrt{\epsilon(\eta) \epsilon(\eta')} G^\text{F}_\text{Q}(x,x') \right] = \delta^n(x-x') \eqend{.}
\end{splitequation}
Comparing with equation~\eqref{propagator_desitter} fulfilled by the de~Sitter propagator we conclude that
\begin{equation}
\label{propagator_q_slowroll_in_ds}
G^\text{F}_\text{Q}(x,x') = \left[ \epsilon(\eta) \epsilon(\eta') \right]^{-\frac{1}{2}} \left[ [ a^{-\epsilon} (1-\epsilon) ](\eta) [ a^{-\epsilon} (1-\epsilon) ](\eta') \right]^\frac{n-2}{2} G^\text{F,dS}_{M^2}(x,x') \eqend{,}
\end{equation}
where the mass $M$ is now given by
\begin{equation}
M^2 = - (n-1) \left( \frac{n-2}{2} \epsilon + \delta \right) H_0^2 \eqend{,}
\end{equation}
and it follows that
\begin{equation}
\label{wightman_q_slowroll}
\tilde{G}^+_\text{Q}(\eta,\eta',\vec{p}) = - \mathi \frac{\pi}{4} \frac{\left[ [ H (1-\epsilon) ](\eta) [ H (1-\epsilon) ](\eta') \right]^\frac{n-2}{2}}{\sqrt{ \epsilon(\eta) \epsilon(\eta') }} (\eta\eta')^\frac{n-1}{2} \hankel1_\nu\left( -\abs{\vec{p}} \eta \right) \hankel2_\nu\left( -\abs{\vec{p}} \eta' \right) \eqend{,}
\end{equation}
where the parameter $\nu$ reads
\begin{equation}
\nu = \frac{n-1}{2} + \frac{n-2}{2} \epsilon + \delta \eqend{.}
\end{equation}
Using Hankel function identities~\cite{dlmf}, we also calculate
\begin{equation}
\label{wightman_dq_slowroll}
\tilde{D}^+_\text{Q}(\eta,\eta',\vec{p}) = \mathi \frac{\pi}{4} \frac{\left[ [ H (1-\epsilon) ](\eta) [ H (1-\epsilon) ](\eta') \right]^\frac{n-2}{2}}{\sqrt{ \epsilon(\eta) \epsilon(\eta') }} (\eta\eta')^\frac{n-1}{2} \hankel1_{\nu-1}\left( -\abs{\vec{p}} \eta \right) \hankel2_{\nu-1}\left( -\abs{\vec{p}} \eta' \right)
\end{equation}
for the function $D^+_\text{Q}$~\eqref{feynman_d_hq}, and in the limit $\delta \to 0$ where $\nu \to \mu$, we recover the small $\epsilon$ approximation to the result~\eqref{wightman_dh_epsconst} of the constant $\epsilon$ case, taking into account the relation~\eqref{propagator_q_epsconst} between $G_\text{Q}$ and $G_\text{H}$ in that case together with the definition~\eqref{feynman_d_hq}.

For the retarded Green's functions, it follows in the same way as for the constant $\epsilon$ case that
\begin{splitequation}
G^\text{ret}_\text{Q}(x,x') &= \mathi \Theta(\eta-\eta') \frac{\left[ [ H (1-\epsilon) ](\eta) [ H (1-\epsilon) ](\eta') \right]^\frac{n-2}{2}}{\sqrt{ \epsilon(\eta) \epsilon(\eta') }} \frac{\Gamma\left( \frac{n-1}{2}+\nu \right) \Gamma\left( \frac{n-1}{2}-\nu \right)}{(4\pi)^\frac{n}{2} \Gamma\left( \frac{n}{2} \right)} \\
&\qquad\times \bigg[ \hypergeom{2}{1}\left[ \frac{n-1}{2}+\nu, \frac{n-1}{2}-\nu; \frac{n}{2}; \frac{1+Z}{2} + \mathi 0 \right] \\
&\qquad\qquad\qquad- \hypergeom{2}{1}\left[ \frac{n-1}{2}+\nu, \frac{n-1}{2}-\nu; \frac{n}{2}; \frac{1+Z}{2} - \mathi 0 \right] \bigg]
\end{splitequation}
and
\begin{splitequation}
D^\text{ret}_\text{Q}(x,x') &= - \mathi \Theta(\eta-\eta') \frac{\left[ [ H (1-\epsilon) ](\eta) [ H (1-\epsilon) ](\eta') \right]^\frac{n-2}{2}}{\sqrt{ \epsilon(\eta) \epsilon(\eta') }} \frac{\Gamma\left( \frac{n-3}{2}+\nu \right) \Gamma\left( \frac{n+1}{2}-\nu \right)}{(4\pi)^\frac{n}{2} \Gamma\left( \frac{n}{2} \right)} \\
&\qquad\times \bigg[ \hypergeom{2}{1}\left[ \frac{n-3}{2}+\nu, \frac{n+1}{2}-\nu; \frac{n}{2}; \frac{1+Z}{2} + \mathi 0 \right] \\
&\qquad\qquad\qquad- \hypergeom{2}{1}\left[ \frac{n-3}{2}+\nu, \frac{n+1}{2}-\nu; \frac{n}{2}; \frac{1+Z}{2} - \mathi 0 \right] \bigg] \eqend{.} \raisetag{2em}
\end{splitequation}
In $n = 4$ dimensions, we can again use the result~\eqref{appendix_hypergeom_jump_sing} to obtain
\begin{splitequation}
G^\text{ret}_\text{Q}(x,x') &= - \Theta(\eta-\eta') \frac{[ H (1-\epsilon) ](\eta) [ H (1-\epsilon) ](\eta')}{4 \pi \sqrt{ \epsilon(\eta) \epsilon(\eta') }} \\
&\quad\times \left[ \Theta(Z-1) \left[ 1 + (\epsilon + \delta) \frac{1+2Z}{1+Z} + (\epsilon + \delta) \ln\left( \frac{1+Z}{2} \right) \right] + \delta(Z-1) \right]
\end{splitequation}
and
\begin{equation}
D^\text{ret}_\text{Q}(x,x') = \Theta(\eta-\eta') \frac{[ H (1-\epsilon) ](\eta) [ H (1-\epsilon) ](\eta')}{4 \pi \sqrt{ \epsilon(\eta) \epsilon(\eta') }} \left[ (\epsilon+\delta) \frac{\Theta(Z-1)}{1+Z} + \delta(Z-1) \right] \eqend{,}
\end{equation}
working to first order in the slow-roll parameters $\epsilon$ and $\delta$ and using that
\begin{equation}
\hypergeom{2}{1}\left( 3 + \alpha, - \alpha; 2; x \right) = 1 + \alpha \left[ \frac{x}{2 (x-1)} + \ln(1-x) \right] + \bigo{\alpha^2} \eqend{,}
\end{equation}
which can be easily derived from the known series expansion of the hypergeometric function~\cite{dlmf}. Again, they are of the general Hadamard form~\cite{hadamard1932,dewittbrehme1960,fullingsweenywald1978,baerginouxpfaeffle2007}.

The other retarded Green's function $G^\text{ret}_\text{H}(x,x')$ is given by the same expression~\eqref{retarded_h_epsconst} as in the constant $\epsilon$ case, with the parameter $\mu$ given by~\eqref{slowroll_mu_def}, and since the relation~\eqref{phpv_commute} is still valid to first order in slow-roll, we also have
\begin{equation}
\label{propagator_2_slowroll}
G^\text{F}_2(x,x') = - \frac{1}{2} \left( \eta \partial_\eta + \eta' \partial_{\eta'} - 2 \mu \right) G^\text{F}_\text{H}(x,x') \eqend{,}
\end{equation}
and the same for the Wightman function.

\subsection{The elliptic condition}
\label{sec_slowroll_elliptic}

Similar to the constant $\epsilon$ case, the would-be ``retarded'' Green's function does not vanish for spacelike separated points, and the non-locality in this gauge is non-causal. Furthermore, the expressions~\eqref{correlator_elliptic_hphi} do not simplify substantially by using the relation~\eqref{slowroll_ha_in_eta} and expanding to first order in the slow-roll parameters, and we thus refrain from writing the resulting correlators explicitly.

\subsection{The hyperbolic condition}
\label{sec_slowroll_hyperbolic}

To simplify the expressions~\eqref{correlator_hyperbolic_hphi}, we use the relation~\eqref{slowroll_ha_in_eta}, the relation~\eqref{propagator_2_slowroll} to express $G^\text{F}_2$ in terms of $G^\text{F}_\text{H}$, and the relations~\eqref{epsconst_df2_dfh_relation} which are still valid to first order in slow-roll. This results in
\begin{equations}[correlator_hyperbolic_hphi_slowroll]
- \mathi \left\langle \mathcal{T} \phi^{(1)}(x) \phi^{(1)}(x') \right\rangle &= - \mathi \left\langle \mathcal{T} \phi^{(1)}(x) h_{\mu\nu}(x') \right\rangle = 0 \eqend{,} \\
- \mathi \left\langle \mathcal{T} h_{00}(x) h_{00}(x') \right\rangle &= (1-2\epsilon) \eta \eta' \partial_\eta \partial_{\eta'} G^\text{F}_\text{Q}(x,x') \eqend{,} \\
- \mathi \left\langle \mathcal{T} h_{00}(x) h_{0k}(x') \right\rangle &= - \frac{\epsilon}{2} \eta \partial_k D^\text{F}_\text{Q}(x,x') - \frac{1}{2} (1-2\epsilon) \eta \eta' \partial_\eta \partial_k G^\text{F}_\text{Q}(x,x') \eqend{,} \\
- \mathi \left\langle \mathcal{T} h_{00}(x) h_{kl}(x') \right\rangle &= \delta_{kl} (1-\epsilon) \eta \partial_\eta G^\text{F}_\text{Q}(x,x') \eqend{,} \\
\begin{split}
- \mathi \left\langle \mathcal{T} h_{0i}(x) h_{0k}(x') \right\rangle &= - \frac{1}{2} \delta_{ik} \left( \eta \partial_\eta + \eta' \partial_{\eta'} - 2 \mu \right) D^\text{F}_\text{H}(x,x') - \frac{n-3}{2 (n-2)} \frac{\partial_i \partial_k}{\laplace} D^\text{F}_\text{H}(x,x') \\
&\quad- \frac{\partial_i \partial_k}{\laplace} \frac{\epsilon \eta' \partial_\eta + \epsilon \eta \partial_{\eta'} + (1-2\epsilon) \eta \eta' \laplace}{4} G^\text{F}_\text{Q}(x,x') \eqend{,}
\end{split} \\
\begin{split}
- \mathi \left\langle \mathcal{T} h_{0i}(x) h_{kl}(x') \right\rangle &= \frac{\delta_{i(k} \partial_{l)}}{\laplace} \partial_\eta \left( \eta \partial_\eta + \eta' \partial_{\eta'} - 2 \mu \right) G^\text{F}_\text{H}(x,x') \\
&\quad- \delta_{kl} \frac{\partial_i}{\laplace} \left[ \frac{1}{n-2} \partial_\eta G^\text{F}_\text{H}(x,x') - \frac{1}{2} \left[ \epsilon \partial_\eta + (1-\epsilon) \eta \laplace \right] G^\text{F}_\text{Q}(x,x') \right] \eqend{,}
\end{split} \\
\begin{split}
- \mathi \left\langle \mathcal{T} h_{ij}(x) h_{kl}(x') \right\rangle &= \left( 2 \delta_{i(k} \delta_{l)j} - \frac{2}{n-2} \delta_{ij} \delta_{kl} \right) G^\text{F}_\text{H}(x,x') + \delta_{ij} \delta_{kl} G^\text{F}_\text{Q}(x,x') \\
&\quad+ 2 \frac{\partial_{(i} \delta_{j)(k} \partial_{l)}}{\laplace} \left( \eta \partial_\eta + \eta' \partial_{\eta'} - 2 \mu \right) G^\text{F}_\text{H}(x,x') \eqend{.}
\end{split}
\end{equations}

As before, the corresponding retarded Green's functions have a proper support in the past light cone. Moreover, one can check that the above expressions agree with the small-$\epsilon$ expansion of the previous result~\eqref{correlator_hyperbolic_hphi_epsconst} for constant $\epsilon$.

\section{Discussion}
\label{sec_discussion}

We have shown how to construct gauge-invariant observables by defining an invariant coordinate system to all orders in perturbation theory around cosmological background spacetimes, given by equations~\eqref{y0_def} and~\eqref{yi_sol}. As with other proposals, these observables are non-local beyond linear order. However, an improvement with respect to a similar proposal by Brunetti et al.~\cite{brunettietal2016} [given by equations~\eqref{x0_def} and~\eqref{xi_sol}] is that the non-locality is causal, i.e., the observables at a point $x$ only depend on the metric and inflaton perturbations in the past light cone of $x$. Therefore, our observables do not suffer from any unphysical ``action-at-a-distance'', and one can apply the rigorous algebraic approach to quantum field theory in curved spacetime~\cite{hollandswald2001,hollandswald2002,hollandswald2005,hollands2008,fredenhagenrejzner2013,hollandswald2015,hack2015,brunettifredenhagenrejzner2016,fredenhagenrejzner2016}.

Unfortunately, the perturbative expansion of these observables quickly becomes unwieldy. Nevertheless, since the observables are explicitly gauge-invariant one can of course calculate their correlation functions in a suitable gauge, which in this case should be adapted to the invariant coordinate system. Namely, the gauge is chosen such that the first-order coordinate corrections~\eqref{x0_sol_firstorder} and~\eqref{yi_sol_firstorder} vanish, which greatly simplifies the perturbative expansion. We have determined the propagators for the metric and inflaton perturbations in this gauge (and also for the proposal by Brunetti et al.), which are given by equations~\eqref{correlator_hyperbolic_hphi} and~\eqref{correlator_hyperbolic_b} [and by~\eqref{correlator_elliptic_hphi} and~\eqref{correlator_elliptic_b} for the Brunetti et al. proposal] for a general cosmological spacetime. These propagators depend on three scalar propagators, whose explicit form depends on the concrete background spacetime. We have also determined these scalar propagators in the two cases most relevant for early universe cosmology:
\begin{itemize}
\item Constant slow-roll parameter $\epsilon$, which covers matter- and radiation-dominated eras. The resulting metric and inflaton propagators are~\eqref{correlator_hyperbolic_hphi_epsconst} for our observables, and~\eqref{correlator_elliptic_hphi_epsconst} for the Brunetti et al. proposal.
\item Slow-roll inflation, where the metric and inflaton propagators are given by~\eqref{correlator_hyperbolic_hphi_slowroll}.
\end{itemize}

With the explicit propagators at hand, calculations of invariant quantum corrections to cosmological observables is now straightforward. One important observable is the local expansion rate, which can be obtained as the divergence of the normalised gradient of the inflaton~\cite{geshnizjanibrandenberger2002} and on the background reduces to the Hubble parameter,
\begin{equation}
H \equiv \frac{\nabla^\mu u_\mu}{n-1} \eqend{,} \qquad u_\mu \equiv \frac{\nabla_\mu \phi}{\sqrt{- \nabla^\mu \phi \nabla_\mu \phi}} \eqend{.}
\end{equation}
Perturbing the metric and inflaton according to equation~\eqref{metric_inflaton_perturbation}, we obtain
\begin{equation}
\tilde{H} = \frac{\tilde{\nabla}^\mu \tilde{u}_\mu}{n-1} = H + \kappa H^{(1)} + \kappa^2 H^{(2)} + \ldots \eqend{,}
\end{equation}
where the first-order correction $H^{(1)}$ reads (in accordance with~\cite{tsamiswoodard2013,miaotsamiswoodard2017})
\begin{equation}
H^{(1)} = - \frac{\laplace \phi^{(1)}}{(n-1) a \phi'} + \frac{H}{2} h_{00} + \frac{1}{2 (n-1) a} \left( \partial_\eta h^k_k - 2 \partial^k h_{0k} \right) \eqend{.}
\end{equation}
The corresponding invariant observable $\mathcal{H} = H + \kappa \mathcal{H}^{(1)} + \kappa^2 \mathcal{H}^{(2)} + \ldots$ is constructed according to~\eqref{scalar_inv_firstorder}, and to first order we obtain [using the expansion of the invariant coordinates~\eqref{x0_sol_firstorder}]
\begin{equation}
\mathcal{H}^{(1)} = H^{(1)} - \tilde{X}^{(\mu)}_{(1)} \partial_\mu H = H^{(1)} + \frac{\epsilon H^2 a}{\phi'} \phi^{(1)} \eqend{.}
\end{equation}
This observable measures the local expansion rate as seen by observers co-moving with the coordinate system $\tilde{X}^{(\mu)}$, in particular [since $\tilde{X}^{(0)} = \eta(\tilde{\phi})$~\eqref{x0_def}] co-moving with the inflaton. As we have seen, the coordinate system $\tilde{X}^{(\mu)}$ involves non-causal non-localities, and one should thus use the coordinate system $\tilde{Y}^{(\mu)}$ instead, but since $\tilde{Y}^{(0)} = \tilde{X}^{(0)}$~\eqref{y0_def}, the difference only shows up at second and higher order. Since $\mathcal{H}$ is invariant, one can use any gauge to calculate its expectation value. However, in the gauge that we have determined in this work where the propagator is given by~\eqref{correlator_hyperbolic_hphi}, the first-order coordinate corrections $\tilde{Y}^{(\mu)}_{(1)}$ vanish, and we have
\begin{equation}
\mathcal{H}^{(1)}_\text{our~gauge} = H^{(1)} \eqend{,} \qquad \mathcal{H}^{(2)}_\text{our~gauge} = H^{(2)} - \tilde{Y}^{(\mu)}_{(2)} \partial_\mu H = H^{(2)} - \kappa^2 \frac{(1-\epsilon+\delta) H}{4 (n-2)} \left[ \phi^{(1)} \right]^2 \eqend{.}
\end{equation}
Moreover, since the gauge condition enforces $\phi^{(1)} = 0$, the last term vanishes and we have even $\mathcal{H}^{(2)}_\text{our~gauge} = H^{(2)}$. This obviously simplifies the calculations, compared to the general result for the second-order correction
\begin{equation}
\mathcal{H}^{(2)} = H^{(2)} - \tilde{Y}_{(1)}^{(\mu)} \partial_\mu H^{(1)} - \left[ \tilde{Y}_{(2)}^{(\mu)} - \tilde{Y}_{(1)}^{(\nu)} \partial_\nu \tilde{Y}_{(1)}^{(\mu)} \right] \partial_\mu H + \frac{1}{2} \tilde{Y}_{(1)}^{(\mu)} \tilde{Y}_{(1)}^{(\nu)} \partial_\mu \partial_\nu H \eqend{.}
\end{equation}

A one-loop calculation of the expectation value of a similar quantity in a de~Sitter background has been completed recently~\cite{miaotsamiswoodard2017}, and we hope to be able to report on the result for the invariant expansion rate $\mathcal{H}$ soon.

\ack
M.~F. thanks Thomas-Paul Hack and Nicola Pinamonti for discussions, and the Centro de Ci{\^e}ncias Naturais e Humanas of the Universidade Federal do ABC for hospitality. W.~L. acknowledges full financial support from Coordena{\c c}{\~a}o de Aperfei{\c c}oamento de Pessoal de N{\'\i}vel Superior (CAPES) through the Graduate Programme in Physics of the Universidade Federal do ABC, and thanks the Maths Department of the University of York for hospitality. We also thank the anonymous referees for helpful comments. This work is part of a project that has received funding from the European Union’s Horizon 2020 research and innovation programme under the Marie Sk{\l}odowska-Curie grant agreement No. 702750 ``QLO-QG''.

\appendix

\section{Metric expansions}
\label{appendix_metric}

Writing a general metric $\tilde{g}_{\mu\nu}$ as background $g_{\mu\nu}$ plus perturbation $h_{\mu\nu}$, we obtain to first order in the perturbation
\begin{equations}[appendix_metric_expansion]
\tilde{g}_{\mu\nu} &= g_{\mu\nu} + \kappa h_{\mu\nu} \eqend{,} \\
\tilde{g}^{\mu\nu} &= g^{\mu\nu} - \kappa h^{\mu\nu} + \bigo{\kappa^2} \eqend{,} \\
\sqrt{-\tilde{g}} &= \sqrt{-g} \left( 1 + \frac{1}{2} \kappa h \right) + \bigo{\kappa^2} \eqend{,} \\
\tilde{\Gamma}^\alpha_{\beta\gamma} &= \Gamma^\alpha_{\beta\gamma} + \frac{1}{2} \kappa \left( \nabla_\beta h^\alpha_\gamma + \nabla_\gamma h^\alpha_\beta - \nabla^\alpha h_{\beta\gamma} \right) + \bigo{\kappa^2} \eqend{,} \\
\begin{split}
\tilde{R}_{\alpha\beta\gamma\delta} &= R_{\alpha\beta\gamma\delta} + \frac{1}{2} \kappa \left( \nabla_\gamma \nabla_{[\beta} h_{\alpha]\delta} - \nabla_\delta \nabla_{[\beta} h_{\alpha]\gamma} + \nabla_\alpha \nabla_{[\delta} h_{\gamma]\beta} - \nabla_\beta \nabla_{[\delta} h_{\gamma]\alpha} \right) \\
&\hspace{4em}- \frac{1}{2} \kappa \left( R_{\alpha\beta\mu[\gamma} h_{\delta]}^\mu + R_{\gamma\delta\mu[\alpha} h_{\beta]}^\mu \right) + \bigo{\kappa^2} \eqend{,}
\end{split} \\
\tilde{R}_{\alpha\beta} &= R_{\alpha\beta} + \kappa \nabla^\delta \nabla_{(\alpha} h_{\beta)\delta} - \frac{1}{2} \kappa \nabla^2 h_{\alpha\beta} - \frac{1}{2} \kappa \nabla_\alpha \nabla_\beta h + \bigo{\kappa^2} \eqend{,} \\
\tilde{R} &= R - \kappa h^{\alpha\beta} R_{\alpha\beta} + \kappa \nabla^\alpha \nabla^\beta h_{\alpha\beta} - \kappa \nabla^2 h + \bigo{\kappa^2} \eqend{.}
\end{equations}
Higher orders can then be obtained by repeating the expansion, i.e.,
\begin{splitequation}
F[\tilde{g}] &= F[g] + \kappa \int \left[ \frac{\delta F[\tilde{g}]}{\delta \tilde{g}_{\mu\nu}(x)} \right]_{\tilde{g} = g} h_{\mu\nu}(x) \sqrt{-g} \total^n x \\
&\quad+ \frac{1}{2} \kappa^2 \iint \left[ \frac{\delta F[\tilde{g}]}{\delta \tilde{g}_{\mu\nu}(x) \delta \tilde{g}_{\rho\sigma}(y)} \right]_{\tilde{g} = g} h_{\mu\nu}(x) h_{\rho\sigma}(y) \sqrt{-g} \total^n x \sqrt{-g} \total^n y + \bigo{\kappa^3} \eqend{,}
\end{splitequation}
and the second functional derivative is calculated by setting $g = \tilde{g}$ after performing the first one, etc.

\section{Some formul{\ae} for propagators and their support}
\label{appendix_support}

We determine some identities for the retarded Green's functions $G^\text{ret}_{\text{H}/\text{Q}/2}$~\eqref{retarded_propagator}. The first two fulfil the differential equation
\begin{equation}
\label{appendix_support_pgret}
P_{\text{H}/\text{Q}} \, G^\text{ret}_{\text{H}/\text{Q}}(x,x') = \delta^n(x-x') \eqend{,}
\end{equation}
where the differential operators $P_{\text{H}/\text{Q}}$ are defined in equation~\eqref{diffops}, while $G^\text{ret}_2$ is a solution of~\eqref{propagator_g2}
\begin{equation}
P_\text{H} \, G^\text{ret}_2(x,x') = P_\text{V} \, G^\text{ret}_\text{H}(x,x') \eqend{,}
\end{equation}
given by
\begin{equation}
\label{appendix_support_g2_sol}
G^\text{ret}_2(x,x') = \int G^\text{ret}_\text{H}(x,y) P_\text{V}(y) \, G^\text{ret}_\text{H}(y,x') \total^n y \eqend{.}
\end{equation}
Defining the commutator
\begin{equation}
\label{appendix_support_commutator}
\Delta(x,x') \equiv G^+(x,x') - G^+(x',x) = - \Delta(x',x) \eqend{,}
\end{equation}
we can write $G^\text{ret}(x,x') = \Theta(\eta-\eta') \Delta(x,x')$. Integrating equation~\eqref{appendix_support_pgret} (with the differential operators acting at $x'$) over $\eta'$ from $\eta - \tau$ to $\eta + \tau$ and taking the limit $\tau \to 0$, we obtain
\begin{splitequation}
\delta^{n-1}(\vec{x}-\vec{x}') &= \lim_{\tau \to 0} \int_{\eta-\tau}^{\eta+\tau} \left[ a^{n-2}(\eta') \laplace G^\text{ret}_\text{H}(x,x') - \partial_{\eta'} \left( a^{n-2}(\eta') \partial_{\eta'} G^\text{ret}_\text{H}(x,x') \right) \right] \total \eta' \\
&= - \lim_{\tau \to 0} \left[ a^{n-2}(\eta') \partial_{\eta'} G^\text{ret}_\text{H}(x,x') \right]_{\eta-\tau}^{\eta+\tau} \\
&= a^{n-2}(\eta) \lim_{\tau \to 0} \left[ - \delta(\eta-\eta') \Delta_\text{H}(x,x') + \Theta(\eta-\eta') \partial_{\eta'} \Delta_\text{H}(x,x')  \right]_{\eta'=\eta-\tau} \\
&= a^{n-2}(\eta) \left. \partial_{\eta'} \Delta_\text{H}(x,x') \right\rvert_{\eta'=\eta} \eqend{,}
\end{splitequation}
because $G^\text{ret}(x,x')$ vanishes for $\eta' = \eta+\tau > \eta$, and $\Delta(x,x')$ is antisymmetric~\eqref{appendix_support_commutator}. In the same way, we obtain
\begin{equations}
\left. \partial_\eta \Delta_\text{H}(x,x') \right\rvert_{\eta'=\eta} &= - a^{2-n}(\eta) \delta^{n-1}(\vec{x}-\vec{x}') \eqend{,} \\
\left. \partial_{\eta'} \Delta_\text{Q}(x,x') \right\rvert_{\eta'=\eta} &= - \left. \partial_\eta \Delta_\text{Q}(x,x') \right\rvert_{\eta'=\eta} = \frac{2 a^{2-n}(\eta)}{(n-2) \epsilon(\eta)} \delta^{n-1}(\vec{x}-\vec{x}') \eqend{.}
\end{equations}

To obtain an expression for the combinations $D^\text{ret}_{\text{H}/\text{Q}}$~\eqref{feynman_d_hq}, we calculate
\begin{splitequation}
\partial_\eta \partial_{\eta'} G^\text{ret}_\text{H}(x,x') &= \partial_\eta \left[ - \delta(\eta-\eta') \Delta_\text{H}(x,x') + \Theta(\eta-\eta') \partial_{\eta'} \Delta_\text{H}(x,x') \right] \\
&= \delta(\eta-\eta') \partial_{\eta'} \Delta_\text{H}(x,x') + \Theta(\eta-\eta') \partial_\eta \partial_{\eta'} \Delta_\text{H}(x,x') \\
&= a^{2-n} \delta^n(x-x') + \Theta(\eta-\eta') \partial_\eta \partial_{\eta'} \Delta_\text{H}(x,x') \eqend{,}
\end{splitequation}
and therefore
\begin{equation}
D^\text{ret}_\text{H}(x,x') = \Theta(\eta-\eta') \frac{\partial_\eta \partial_{\eta'}}{\laplace} \Delta_\text{H}(x,x') \eqend{.}
\end{equation}
In the same way, it follows that
\begin{equation}
D^\text{ret}_\text{Q}(x,x') = \Theta(\eta-\eta') \frac{\partial_\eta \partial_{\eta'}}{\laplace} \Delta_\text{Q}(x,x') \eqend{.}
\end{equation}
For a single time derivative, we calculate in general
\begin{equation}
\partial_\eta G^\text{ret}(x,x') = \delta(\eta-\eta') \Delta(x,x') + \Theta(\eta-\eta') \partial_\eta \Delta(x,x') = \Theta(\eta-\eta') \partial_\eta \Delta(x,x') \eqend{,}
\end{equation}
such that such terms do not generate Dirac $\delta$'s. Lastly, we calculate for the retarded combination $D^\text{ret}_2$~\eqref{feynman_d_2}
\begin{splitequation}
D^\text{ret}_2(x,x') &= \frac{\partial_\eta \partial_{\eta'}}{\laplace} G^\text{ret}_2(x,x') = \frac{1}{\laplace} \int \partial_\eta G^\text{ret}_\text{H}(x,y) P_\text{V}(y) \, \partial_{\eta'} G^\text{ret}_\text{H}(y,x') \total^n y \\
&= \frac{1}{\laplace} \int \Theta(\eta-y^0) \partial_\eta \Delta_\text{H}(x,y) P_\text{V}(y) \left[ \Theta(y^0-\eta') \partial_{\eta'} \Delta_\text{H}(y,x') \right] \total^n y
\end{splitequation}
using the solution~\eqref{appendix_support_g2_sol}, and since $P_\text{V} = a^{n-2} \laplace$~\eqref{diffops}, no extra Dirac $\delta$'s are generated in this case as well.

Going to (spatial) Fourier space, one then can check in each case that the time derivatives generate each an extra factor of $\abs{\vec{p}}$, which cancel the factor $\abs{\vec{p}}^{-2}$ coming from the inverse Laplacian. The same is true for the projection operators $\partial_i \partial_j \laplace^{-1}$, and thus the small-$\vec{p}$ behaviour of the retarded Green's functions in Fourier space is unchanged. The problematic IR divergences for the two-point function, which arise from the small-$\vec{p}$ behaviour and lead to terms which are supported on equal-time hypersurfaces, are state-dependent and cancel out in the retarded Green's functions. Since the small-$\vec{p}$ behaviour is unchanged when acting with time derivatives over inverse Laplacians and projection operators, no new such terms are generated, and the retarded Green's functions keep their proper support inside the past light cone. This can also be verified explicitly, by acting with the inverse Laplacians on the Fourier-transformed retarded Green's function and verifying that the small-$\vec{p}$ divergences cancel and the integral~\eqref{fourier_integral_i} can be used, see for example equation~\eqref{wightman_dh_epsconst}.

\section{Hypergeometric function across the branch cut}
\label{appendix_hypergeom}

We use the Euler integral formula~\cite{dlmf} for the Gau{\ss} hypergeometric function
\begin{equation}
\hypergeom{2}{1}\left( a,b; c; z \right) = \frac{\Gamma(c)}{\Gamma(b) \Gamma(c-b)} \int_0^1 t^{b-1} (1-t)^{c-b-1} (1-zt)^{-a} \total t \eqend{,}
\end{equation}
valid for $\Re c > \Re b > 0$ and all complex $z \not\in [1,\infty)$. The standard result for the discontinuity of $\hypergeom{2}{1}$ across the branch cut, which we denote by
\begin{equation}
\Delta \hypergeom{2}{1}\left( a,b; c; z \right) \equiv \hypergeom{2}{1}\left( a,b; c; z+\mathi 0 \right) - \hypergeom{2}{1}\left( a,b; c; z-\mathi 0 \right) \eqend{,}
\end{equation}
proceeds as follows: assume that $z > 1$. Since
\begin{equation}
(1-zt)^{-a} = \exp\left[ - a \left( \ln\abs{1-zt} + \mathi \arg(1-zt) \right) \right] \eqend{,}
\end{equation}
and for $x \in \mathbb{R}$
\begin{equation}
\arg(x + \mathi 0) = \pi \, \Theta(-x) \eqend{,} \qquad \arg(x - \mathi 0) = - \pi \, \Theta(-x) \eqend{,}
\end{equation}
we have for all $t \geq 0$
\begin{equation}
\Delta (1-zt)^{-a} = [ 1 - (z+\mathi 0) t ]^{-a} - [ 1 - (z-\mathi 0) t ]^{-a} = \abs{1-zt}^{-a} \, \Theta(zt-1) \, 2 \mathi \sin(a \pi)
\end{equation}
and thus
\begin{equation}
\Delta \hypergeom{2}{1}\left( a,b; c; z \right) = \frac{2 \mathi \sin(a \pi) \Gamma(c)}{\Gamma(b) \Gamma(c-b)} \int_\frac{1}{z}^1 t^{b-1} (1-t)^{c-b-1} (zt-1)^{-a} \total t \eqend{.}
\end{equation}
We now perform a change of variables
\begin{equation}
t = \frac{1}{z} + \frac{z-1}{z} s
\end{equation}
to obtain
\begin{splitequation}
\Delta \hypergeom{2}{1}\left( a,b; c; z \right) &= \frac{2 \mathi \sin(a \pi) \Gamma(c)}{\Gamma(b) \Gamma(c-b)} (z-1)^{c-b-a} z^{1-c} \int_0^1 s^{-a} (1-s)^{c-b-1} [ 1 + (z-1) s ]^{b-1} \total s \\
&= \frac{2 \mathi \sin(a \pi) \Gamma(1-a) \Gamma(c)}{\Gamma(b) \Gamma(c-b-a+1)} (z-1)^{c-b-a} z^{1-c} \\
&\qquad\times \hypergeom{2}{1}\left( 1-b,1-a; c-b-a+1; 1-z \right) \\
&= \frac{2 \pi \mathi \, \Gamma(c) (z-1)^{c-b-a} z^{1-c}}{\Gamma(a) \Gamma(b) \Gamma(c-b-a+1)} \hypergeom{2}{1}\left( 1-b,1-a; c-b-a+1; 1-z \right) \eqend{,}
\end{splitequation}
where in the second step we used the integral representation again, and the third step follows from the $\Gamma$ reflection identity. To bring this into the usual form, we use the Euler transformation~\cite{dlmf} (which follows from applying the Pfaff transformation twice, which in turn is obtained by changing $t \to (1-t)$ in the integral representation) and get
\begin{equation}
\label{appendix_hypergeom_jump_reg}
\Delta \hypergeom{2}{1}\left( a,b; c; z \right) = \frac{2 \pi \mathi \, \Gamma(c) (z-1)^{c-b-a}}{\Gamma(a) \Gamma(b) \Gamma(c-b-a+1)} \hypergeom{2}{1}\left( c-a, c-b; c-b-a+1; 1-z \right) \eqend{.}
\end{equation}

However, this derivation is obviously problematic when $z = 1$ and $\Re (c-a-b) \leq -1$, which happens for large enough $\Re a$ even if $\Re (c-b) > 0$. In this case, the integral representation has a non-integrable singularity at $z = 1$, and from the Euler transformation we know that the hypergeometric function then diverges as $(1-z)^{c-a-b}$. To cover this case, we use the integral representation for the transformed hypergeometric function:
\begin{splitequation}
\hypergeom{2}{1}\left( a,b; c; z \right) &= (1-z)^{c-a-b} \hypergeom{2}{1}\left( c-a, c-b; c; z \right) \\
&= (1-z)^{c-a-b} \frac{\Gamma(c)}{\Gamma(b) \Gamma(c-b)} \int_0^1 t^{c-b-1} (1-t)^{b-1} (1-zt)^{a-c} \total t \eqend{.}
\end{splitequation}
We can now calculate the discontinuity by first subtracting sufficiently many terms of the Taylor expansion of $(1-zt)^{a-c}$ around $z = 1$, and treat those terms separately. The separate terms will give a discontinuity only if $c-a-b \in \mathbb{Z}$, and for the moment let us treat the case $c-a-b=-1$ (and $\Re a > 1$, $\Re b > 0$). Then
\begin{splitequation}
&\hypergeom{2}{1}\left( a, b; a+b-1; z \right) = (1-z)^{-1} \frac{\Gamma(a+b-1)}{\Gamma(b) \Gamma(a-1)} \int_0^1 t^{a-2} \left( \frac{1-t}{1-zt} \right)^{b-1} \total t \\
&\quad= (1-z)^{-1} \frac{\Gamma(a+b-1)}{\Gamma(b) \Gamma(a-1)} \int_0^1 t^{a-2} \left[ \left( \frac{1-t}{1-zt} \right)^{b-1} - 1 \right] \total t + (1-z)^{-1} \frac{\Gamma(a+b-1)}{\Gamma(b) \Gamma(a)} \eqend{.}
\end{splitequation}
We now recall the formul{\ae}
\begin{equation}
\ln(x + \mathi 0) = \ln\abs{x} + \mathi \pi \Theta(-x) \eqend{,} \qquad \frac{1}{x + \mathi 0} = \mathcal{P}\!f \frac{1}{x} - \mathi \pi \delta(x) \eqend{,}
\end{equation}
from which it follows by repeated differentiation that
\begin{equation}
\frac{1}{( x + \mathi 0 )^k} - \frac{1}{( x - \mathi 0 )^k} = \frac{(-1)^k}{(k-1)!} 2 \mathi \pi \delta^{(k-1)}(x) \eqend{.}
\end{equation}
We furthermore calculate for $z > 1$
\begin{equation}
\Delta \left[ (1-z)^{-1} \left[ \left( \frac{1-t}{1-zt} \right)^{b-1} - 1 \right] \right] = \Theta(zt-1) \abs{z-1}^{-1} (1-t)^{b-1} (zt-1)^{1-b} \, 2 \mathi \sin(b \pi)
\end{equation}
and thus
\begin{splitequation}
\Delta \hypergeom{2}{1}\left( a, b; a+b-1; z \right) &= \frac{2 \mathi \sin(b \pi) \Gamma(a+b-1)}{\Gamma(b) \Gamma(a-1) (z-1)} \Theta(z-1) \int_\frac{1}{z}^1 t^{a-2} (1-t)^{b-1} (zt-1)^{1-b} \total t \\
&\qquad+ 2 \mathi \pi \delta(1-z) \frac{\Gamma(a+b-1)}{\Gamma(b) \Gamma(a)} \\
&= - \frac{2 \mathi \pi \Gamma(a+b-1)}{\Gamma(b-1) \Gamma(a-1)} \Theta(z-1) z^{2-a-b} \hypergeom{2}{1}\left( 2-a, 2-b; 2; 1-z \right) \\
&\qquad+ 2 \mathi \pi \delta(1-z) \frac{\Gamma(a+b-1)}{\Gamma(b) \Gamma(a)} \eqend{,} \raisetag{2em}
\end{splitequation}
using the same variable change $t = [ 1 + (z-1) s ]/z$ as before. We can use again the Euler transformation to bring it into a nicer form, which is
\begin{splitequation}
\Delta \hypergeom{2}{1}\left( a, b; a+b-1; z \right) &= - \frac{2 \pi \mathi \, \Gamma(a+b-1)}{\Gamma(a-1) \Gamma(b-1)} \\
&\qquad\times \left[ \Theta(z-1) \hypergeom{2}{1}\left( a, b; 2; 1-z \right) - \frac{\delta(1-z)}{(a-1) (b-1)} \right] \eqend{.}
\end{splitequation}
Note that the first term is the limit of the general result as $c \to a+b-1$, but the usual treatment misses the $\delta$ term.

For $c-a-b = -k$ with $k \in \mathbb{N}$ we can thus just calculate the singular terms, and take the regular one from the previous result. We have
\begin{splitequation}
&\hypergeom{2}{1}\left( a, b; a+b-k; z \right) = (1-z)^{-k} \frac{\Gamma(a+b-k)}{\Gamma(b) \Gamma(a-k)} \int_0^1 t^{a-k-1} (1-t)^{b-1} (1-zt)^{-b+k} \total t \\
&\quad= (1-z)^{-k} \frac{\Gamma(a+b-k)}{\Gamma(b) \Gamma(a-k)} \sum_{\ell=0}^{k-1} \frac{\Gamma(b-k+\ell)}{\Gamma(b-k) \ell!} (z-1)^\ell \int_0^1 t^{a-k-1+\ell} (1-t)^{k-1-\ell} \total t + \text{reg.} \eqend{,}
\end{splitequation}
where ``reg.'' means terms whose discontinuity is regular as $z \to 1$. The integral can be done and we obtain
\begin{splitequation}
&\Delta \hypergeom{2}{1}\left( a, b; a+b-k; z \right) = \frac{\Gamma(a+b-k)}{\Gamma(a) \Gamma(b)} \sum_{\ell=0}^{k-1} \frac{\Gamma(a-k+\ell) \Gamma(b-k+\ell)}{\Gamma(a-k) \Gamma(b-k) \ell!} (-1)^\ell 2 \pi \mathi \delta^{(k-1-\ell)}(z-1) \\
&\hspace{4em}+ \lim_{\ell \to k} \frac{2 \pi \mathi \, \Gamma(a+b-\ell)}{\Gamma(a) \Gamma(b) \Gamma(1-\ell)} \Theta(z-1) (z-1)^{-\ell} \hypergeom{2}{1}\left( b-\ell, a-\ell; 1-\ell; 1-z \right) \eqend{,}
\end{splitequation}
and using hypergeometric identities~\cite{dlmf} to evaluate the limit we obtain finally
\begin{splitequation}
\label{appendix_hypergeom_jump_sing}
&\Delta \hypergeom{2}{1}\left( a, b; a+b-k; z \right) = (-1)^k \frac{2 \pi \mathi \, \Gamma(a+b-k)}{\Gamma(a-k) \Gamma(b-k) k!} \\
&\quad\times \left[ \Theta(z-1) \hypergeom{2}{1}\left( a, b; k+1; 1-z \right) - \sum_{\ell=0}^{k-1} \frac{\Gamma(a-1-\ell) \Gamma(b-1-\ell) k!}{\Gamma(a) \Gamma(b) \Gamma(k-\ell)} (-1)^\ell \delta^{(\ell)}(z-1) \right] \eqend{.}
\end{splitequation}

% the iopart definition
\providecommand\newblock{\ }
\bibliography{literature}

\end{document}